\documentclass[12pt,oneside,reqno]{amsart}
\usepackage{graphicx}
\usepackage{cancel}
\pdfoutput=1
\usepackage[margin=1in]{geometry}
\usepackage{abstract}
\usepackage{amssymb, amsmath, amsthm, mathrsfs, amscd, amsaddr}
\usepackage{subcaption, color, graphicx}
\usepackage[dvipsnames]{xcolor}
\usepackage{microtype, siunitx, fancyhdr, polynom}
\usepackage{tikz, tikz-cd, physics, multicol, multirow}
\usepackage[all, cmtip]{xy}
\usepackage{comment}
\usepackage{float}
\usepackage{placeins}

\usepackage[colorlinks,urlcolor=blue,citecolor=blue]{hyperref}

\def\hhref#1{\href{http://arxiv.org/abs/#1}{arXiv:#1}}

\newcommand\qt{\widetilde{q}}
\def\ff#1#2{{\textstyle\frac{#1}{#2}}}

 \usepackage{stackrel,nicefrac}
 \usepackage{mathtools}

\usetikzlibrary{decorations.markings}

\allowdisplaybreaks

\tikzset{->-/.style={decoration={
  markings,
  mark=at position #1 with {\arrow{>}}},postaction={decorate}}}

\numberwithin{equation}{section}

\def\tilde{\widetilde}

\begin{document}

    \vspace*{-1cm}%
    \begin{minipage}[t]{16cm}
      \begin{flushright}
       BONN-MPIM-2024;  BONN-TH-2025-14\\
      \end{flushright}
    \end{minipage}%
  \vspace*{1cm}

\title{$c_{\rm eff}$ from Resurgence at the Stokes Line}

\author{Griffen Adams$^1$, Ovidiu Costin$^2$, Gerald V. Dunne$^1$, Sergei Gukov$^{3}$, Oğuz \"Oner$^4$  }

\address{$^1$Physics Department, University of Connecticut, 196 Auditorium Road,
Storrs, CT 06269, USA}
\address{$^2$Mathematics Department, The Ohio State University, 231 W. 18th Avenue,  Columbus, Ohio 43210, USA}
\address{$^3$Division of Physics, Mathematics and Astronomy, California Institute of Technology, 1200~E.~California Blvd., Pasadena, CA 91125, USA}
\address{$^4$Bethe Center for Theoretical Physics, Universit\"at Bonn, D-53115, Germany}

\setcounter{tocdepth}{2}
\email{griffen.adams@uconn.edu}
\email{costin@math.osu.edu}
\email{gerald.dunne@uconn.edu}
\email{gukov@math.caltech.edu}
\email{ooner@caltech.edu}

\maketitle

\begin{abstract}

In recent papers \cite{CDGG,ACDGO}, a new method to cross the natural boundary has been proposed, and applied to Mordell-Borel integrals arising in the study of Chern-Simons theory, based on decompositions into {\it resurgent cyclic orbits}. Resurgent analysis on the Stokes line leads to a unique transseries decomposition in terms of unary false theta functions, which can be continued across the natural boundary to produce dual $q$-series whose integer-valued coefficients enumerate BPS states. This constitutes a deeper new manifestation of resurgence in quantum field theoretic path integrals. In this paper we show that the algebraic structure of the {\it resurgent cyclic orbits}, combined with just the leading term of the $q$-series, completely determines the large order rate of growth of the dual $q$-series coefficients. The essential exponent of this asymptotic growth has a Cardy-like interpretation \cite{GJ} of an effective central charge in a 3 dimensional quantum field theory with $\mathcal{N}=2$ supersymmetry related to the Chern-Simons theory through the $3d$-$3d$ correspondence.

\end{abstract}

\setcounter{tocdepth}{2}
\tableofcontents

\section{Introduction and Motivation}
In recent papers \cite{CDGG,ACDGO} it has been shown using resurgent analysis \cite{sauzin,Co08,Marino:2012zq,Dorigoni:2014hea,ABS19,CD20b} that the Mordell-Borel integrals arising in the analysis of non-perturbative completions of Chern-Simons theories on certain three-dimensional manifolds, have a unique transseries decomposition into real and imaginary parts on the Stokes line. Furthermore, these Mordell-Borel integrals decompose into natural {\it resurgent cyclic orbits} ({\it resurgent orbits}, for short) under the $SL(2, \mathbb Z)$ operations which probe the natural boundary.

Resurgent analysis  predicts that the algebraic structure of these orbits should be preserved as the integrals cross the natural boundary, and this perspective provides an efficient numerical procedure to determine the integer-valued coefficients of the $q$-series expansions on {\it the other side} of the natural boundary. These integer coefficients enumerate BPS states in three-dimensional $\mathcal N=2$ supersymmetric QFTs via the 3d-3d correspondence or, equivalently, in an M-theory setup that involves fivebranes wrapped on Lagrangian submanifolds in a local Calabi-Yau 3-fold. The asymptotic growth of the BPS state count in such systems defines a quantity $c_{\rm eff}$ analogous to the black hole entropy in gravitational theories \cite{SV}:
\begin{eqnarray}
    b_n \sim \text{Re} \Big[ \exp \left( \sqrt{ \frac{2 \pi^2}{3} c_{\rm eff}\, n} + 2 \pi i rn \right) \Big]
     \qquad, \quad n\to \infty
     \label{Cardygeneral}
\end{eqnarray}
As emphasized in \cite{GJ}, in fairly general contexts of BPS state counting, complex Chern-Simons theory, logarithmic vertex algebras or other problems\footnote{See e.g. \cite{Watson,Dra,ono,Zwe08,Zag09,GM12,Bringmann,Bring15,DMZ12,CD20c} for number theoretic contexts, \cite{Guk05,GPV,GPPV,CGPS,EGGKPS,CGP,CS} for relevant topological aspects, and \cite{GMP,Chun17,CCFGH,CFG19,CCKPG,Park21,Wu20,GM,AM22,Garoufalidis:2021lcp,HLSS22,Fantini:2024ihf,Gukov:2024vbr} for applications of resurgent analysis to complex Chern-Simons theory.} where generating $q$-series $\chi (q) = \sum_n b_n q^{\Delta + n}$ naturally appear, there can be some crucial differences from the familiar Cardy behavior \cite{Cardy}. First, in general the ``effective central charge''  can be complex-valued, $c_{\rm eff} \in \mathbb{C}$. Hence, the relation to integer coefficients $b_n \in \mathbb{Z}$ can only be meaningful if we use the real part $\text{Re} (\ldots)$ in \eqref{Cardygeneral}. This feature leads to an important consequence: the coefficients $b_n$ can exhibit oscillations of growing amplitude. In other words, in general, the Cardy growth controls the growth of the amplitude.

The second new feature of \eqref{Cardygeneral} is the appearance of a rational number $r \in \mathbb{Q} / \mathbb{Z}$ that is determined by the position, $q_* = e^{2\pi i nr}$, of the dominant saddle near the unit circle on the $q$-plane. In turn, the denominator of the rational number $r$ determines the number of different {\it branches} \cite{GJ}. In other words, in their asymptotic behavior the collection of coefficients $\{ b_n \}$ breaks into a finite set of groups and the number of such ``packets'' is precisely the denominator of $r$. In simple examples, the denominator of $r$ is 1 and it can be ignored since there is only one {\it branch}.

The quantity $c_{\rm eff}$ that will be of our main interest in this paper provides valuable information for both sides of 3d-3d correspondence. On the 3-manifold side, it is an invariant of a 3-manifold. And, the main question is: What invariant is it? From the preliminary work \cite{ACDGO,GJ} it is clear that $c_{\rm eff}$ is essentially an integral $\text{CS} (\alpha_*)$, where $\alpha_*$ is one of complex flat connections on $M_3$ or, possibly, a non-flat connection \cite{Gukov:2024vbr}. When $\alpha_*$ is a complex flat connection, a natural question is: Which one? Is there a topological characterization of $\alpha_*$? And, when it is not a flat connection, the characterization of $\alpha_*$ is even more interesting. Another important question is to understand the integral lift of $\text{CS} (\alpha_*)$ that determines $c_{\rm eff}$; it is itself an integer invariant of $M_3$ and the analogous question here is: What integer invariant is it? Is it a new invariant, or a combination of known ones?

The quantity $c_{\rm eff}$ is equally interesting on the ``supersymmetric side'' of the 3d-3d correspondence. In the 3d $\mathcal N=2$ theory, it provides a candidate for the $c$-function or $a$-function, analogous to the familiar gravitational anomaly coefficients in two or four dimensions, cf. \cite{Zamolodchikov:1986gt,Cardy:1988cwa,Barnes:2004jj,Komargodski:2011vj,Jafferis:2011zi,Klebanov:2011td,Casini:2012ei,Klebanov:2012va}. Since at present the majority of theories $T[M_3]$ do not admit a Lagrangian description, and there is some evidence that such a description may not exist \cite{Chung:2019khu,Chung:2023qth}, the quantity $c_{\rm eff}$ becomes a valuable measure of degrees of freedom in such theories. In order to understand its behavior under RG flows, as well as to answer the above questions on the other side of the 3d-3d correspondence, we need better ways to calculate $c_{\rm eff}$ and to explore its origin in the Borel plane. That is precisely the main goal of the present work.

Specifically, here we focus on the implications of the principle of preservation of relations \cite{sauzin}, which has been argued in \cite{CDGG,ACDGO} to be a powerful new approach to the crossing of the natural boundary.
We use the results of \cite{CDGG,ACDGO} to study the large-order growth of the BPS degeneracies. The large order growth has the leading asymptotic form \eqref{Cardygeneral} with $c_{\rm eff}$ having the interpretation of an effective central charge (suitably normalized), which measures the density of BPS states, leading to a Cardy-like formula  \cite{GJ}. We show here that the physical parameter $c_{\rm eff}$ is essentially fixed by the algebraic structure that is determined on the Stokes line, just requiring one more piece of information, the leading nonzero term in the $q$-series, which is obtained numerically using the method introduced in \cite{ACDGO}.

The $3d$ $\mathcal{N}=2$ superconformal field theory $T[M_3]$ has an associated superconformal index \cite{GJ}
\begin{equation}
\label{eq:index}
\mathcal{I}(q)=\frac{1}{\eta(q)^2}\sum_a\left|\mathcal{W}_a\right|  \widehat{Z}_a\left(M_3, q\right) \widehat{Z}_a\left(M_3, q^{-1}\right) \quad \in \mathbb{Z}[[q]],
\end{equation}
where $\left|\mathcal{W}_a\right|$ are symmetry factors, and $\eta(q)$ is a normalization factor. This has an associated effective central charge related to the large order growth of its $q$-series coefficients. The large order growth can therefore be deduced from the growth of the coefficients of the $q$-series $\widehat{Z}(M_3,q)$ and $\widehat{Z}(M_3,q^{-1}) = \widehat{Z}(\overline{M_3},q)$, the latter referring to the orientation-reversed 3-manifold $\overline{M_3}$. The effective central charge of the superconformal index receives contributions from each $q$-series, which are referred to as the half indices $c_{\text{eff},\frac{1}{2}}$ of the theories $T[M_3]$ and $T[\overline{M_3}]$. As discussed in \cite{GJ}, when $M_3$ is a Brieskorn sphere, $c_{\text{eff},\frac{1}{2}}^{T[M_3]}$ can be computed directly from perturbative information. However, to compute $c_{\text{eff},\frac{1}{2}}^{T[\overline{M_3}]}$ we need the full transseries representation of the partition function, where we find that a particular non-abelian flat connection contributes to the central charge. Using the numerical methods proposed in \cite{ACDGO}, we can find this connection and its associated Chern-Simons value to compute the effective central charge of $T[\overline{M_3}]$. 

In this paper, we discuss examples for 3-manifolds that are accessible to the resurgent approach as well as other approaches. For example, another approach to compute $c_{\rm eff}$ has been proposed in a related paper \cite{mrunmay}, based on a regularized positive surgery formula \cite{Park21}, together with modularity arguments.
Using mixed mock modularity of $\widehat{Z}$ invariants, \cite{mrunmay} finds that the regularized positive surgery formula in \cite{Park21} is not generally consistent with the form of $c_{\rm eff}$  proposed in \cite{GJ}.
In our resurgent approach we do not use the regularized positive surgery formula in \cite{Park21}, and our results have the expected structural form for BPS spectra of Brieskorn spheres.
We further present results for other Brieskorn spheres that are currently not accessible with alternative methods. An interesting feature of these computations is that very different approaches lead to agreements in some nontrivial examples, while the incompatibilities in other examples suggest the existence of further deeper structure of the problem. This presents an exciting discovery potential for future work on this topic.

In Section 2 we explain how the algebraic structure determines the growth rate of the integer-valued coefficients for certain $q$-series. In Section 3 we define the Mordell-Borel integrals associated with the duals of false theta functions, and with the linear combinations of four false theta functions associated with Brieskorn spheres, based on \cite{ACDGO}. Section 4 contains our detailed results for the duals of false theta functions, while Section 5 describes the duals relevant for Brieskorn spheres. Sections 4 and 5 contain comparisons with results from other proposed approaches to crossing the natural boundary.

\section{Legendre Transform Duality Between Asymptotics and Large Order Growth}
\label{sec:legendre}

Our analysis is based on the following simple duality between the large order growth of the coefficients $b_n$ of a $q$-series
\begin{eqnarray}
    Y(q):=
    \sum_{n=0}^\infty b_n\, q^n
    \qquad, \qquad q:=e^{-t},
    \label{eq:yq1}
\end{eqnarray} 
and the asymptotics of $Y(e^{-t})$ and  $Y(e^{-\pi^2/t})$. 
\begin{eqnarray}
  {\rm asymptotics\, in\,} t \qquad  \Longleftrightarrow\qquad  {\rm large\, order\, growth\, of\,} b_n
\end{eqnarray}
We further assume for simplicity that all the $b_n$ coefficients have the same sign as $n\to\infty$.  The argument is similar if the  $b_n$ alternate in sign. All the physical examples here and in \cite{ACDGO} have either a same-sign or alternating-sign pattern. This has the important consequence that the most sensitive growth condition can be obtained from one of the directions approaching $q\to 1$ or $q\to -1$, whichever produces the non-alternating sign pattern. 

In the study of Mordell-Borel integrals \cite{ACDGO} one encounters expressions decomposing a function $H(t)$ into a $q$-series $Y_1(q)$ and another $\qt$-series $Y_2(\qt)$ in terms of $\qt=e^{-\pi^2/t}$:
\begin{eqnarray}
    H(t) = q^{-\alpha_1} \,Y_1(q)+\sqrt{\frac{\pi}{t}}\, 
    \qt^{\, -\alpha_2}\, Y_2(\qt).
    \label{eq:hyy}
\end{eqnarray}
This expression is valid for $t>0$, with $\alpha_1>0$, $\alpha_2>0$,\footnote{The special {\it self-dual} case $Y_1=Y_2$, and $\alpha_1=\alpha_2$, occurs frequently in applications. See \cite{ACDGO}.} and with the following $t\to 0^+$ limiting behavior:
\begin{eqnarray}
    H(t)\to A \quad {\rm as} \quad t\to 0^+\quad ; \qquad Y_2(\qt)\to B\quad {\rm as} \quad \qt\to 0^+,
\end{eqnarray}
where $A$ and $B$ are nonzero constants. Since $H(t)$ is finite in this limit, consistency implies exponential growth of $Y_1(q)$  as $q\to 1^-$:
\begin{eqnarray}
    Y_1(e^{-t})\sim - B\, \sqrt{\frac{\pi}{t}}\, \exp\left[\frac{\alpha_2\,  \pi^2}{t}\right]+A+\dots \qquad, \quad t\to 0^+.
    \label{eq:y-asym}
\end{eqnarray}
This asymptotic information is sufficient to determine the leading large order asymptotics of the coefficients $b_n$ of the $q$-series $Y_1(q)$. The correspondence is
\begin{eqnarray}
   Y_1(e^{-t})&\sim& - B\, \sqrt{\frac{\pi}{t}}\, \exp\left[\frac{c\,  \pi^2}{t}\right]+\dots \qquad  t\to 0^+,
   \nonumber\\
   & \updownarrow& 
   \label{eq:corr}\\
  b_n&\sim& -\frac{B}{2}\, n^{-1/2} \, e^{2\pi\sqrt{c\, n}} +\dots 
  \qquad n\to +\infty. 
  \nonumber
\end{eqnarray}
This follows from a saddle-point analysis of the sum in \eqref{eq:yq1}. Write $b_n := e^{F(n)}$, where we assume that $F(n)$ varies smoothly for large $n$. Then the sum can be approximated by a Gaussian integral, yielding:
\begin{eqnarray}
    Y(q) \approx \sqrt{\frac{\pi}{F''(n_{crit})}} e^{\max_n[-nt + F(n)]}.
    \label{eq:yq2}
\end{eqnarray}
In the exponent, we recognize a Legendre transform: \(f^*(x^*) = \sup_x [x^*x - f(x)]\), with \(x^* = -t\), \(x=n\) and \(f = -F\). Since the Legendre transform is an involution, we can deduce the form of $F(n)$ from the behavior of $Y(e^{-t})$ in the limit $t\to 0^+$. 

Explicitly, the asymptotics in \eqref{eq:y-asym} suggests considering large-order growth of the expansion coefficients of the form 
\begin{eqnarray}
    b_n\sim D\, n^{d} \, e^{2\pi\sqrt{c\, n}}\qquad, \quad n\to +\infty.
    \label{eq:growth-form}
\end{eqnarray}
The constants $D$, $d$ and $c$ are to be determined.
The critical point of the summand in \eqref{eq:yq1} is
\begin{eqnarray}
    n_{crit}=\frac{1}{2t^2}\left(c\, \pi^2+2d \,t+\pi\sqrt{c^2\pi^2+4c\, d\, t}\right).
\end{eqnarray}
Approximating the sum as a Gaussian integral around this saddle point, we obtain the leading estimate
\begin{eqnarray}
    Y(q)\sim 2D \left(\frac{c\,\pi^2}{t^2}\right)^{d+\frac{1}{2}}\sqrt{\frac{\pi}{t}} \, e^{c\,\pi^2/t} +\dots \qquad , \quad t\to 0^+.
    \label{eq:gaussian}
\end{eqnarray}
Comparing with \eqref{eq:y-asym} fixes the choices: $c=\alpha_2$, $d=-\frac{1}{2}$ and $D=-B/2$. The fact that $d=-\frac{1}{2}$ in \eqref{eq:growth-form} follows from the universal $\sqrt{\frac{\pi}{t}}$ factor in the decomposition identity in \eqref{eq:hyy}.
A straightforward extension of this argument predicts subleading corrections to the large $n$ growth, used below in Section \ref{sec:sub}.

We can also extend these results to the case where the decomposition \eqref{eq:hyy} is a vector of such identities.
See Sections \ref{sec:false}-\ref{sec:brieskorn} below. This is the basic structure underlying the transformation properties of false and mock theta functions. 
In \cite{CDGG,ACDGO} it has been shown that classes of Mordell-Borel integrals decompose into $SL(2, \mathbb Z)$ orbits, which we refer to as {\it resurgent cyclic orbits}, and which arise in the analysis of Chern-Simons theory on certain 3 dimensional manifolds. These orbits satisfy identities with structure analogous to \eqref{eq:hyy}. Furthermore, a numerical algebraic method was introduced to generate the integer-valued coefficients $b_n$ of the corresponding dual $q$-series. Here we show that the resurgent cyclic orbit structure encodes the growth rate of the $q$-series coefficients, and show that the leading analytic asymptotics matches very accurately the numerical results in \cite{CDGG,ACDGO}. This argument generalizes straightforwardly to include corrections beyond the leading asymptotic growth, as is outlined in the Appendix \ref{sec:sub}.

\section{Mordell-Borel Integrals and False Theta Functions on the Stokes Line}
\label{sec:mordell-borel}

We first recall some basic notation, definitions and results from \cite{CDGG,ACDGO}.
We define the following Mordell-Borel integrals as building blocks for our analysis \cite{GM12}:
\begin{eqnarray}
JS_{(p,a)}(t)&:=&\frac{1}{t} \int_0^\infty du\, e^{-p u^2/t} \, \frac{\sinh[(p-a)u]}{\sinh[p u]}, 
\label{eq:js}\\
JC_{(p,a)}(t)&:=&\frac{1}{t} \int_0^\infty du\, e^{-p u^2/t} \, \frac{\cosh[(p-a)u]}{\cosh[p u]} 
\label{eq:jc}.
\end{eqnarray}
Initially defined in \cite{GM12} for $t>0$, we will analyze their continuation to the complex $t$ plane. 
In this paper we restrict our attention to $p$ and $a$ positive integers with $0 < a < p$, and with $p$ odd. 

In \cite{CDGG,ACDGO} it is shown that both the $t\to 0^+$ and $t\to +\infty$ expansions of $JS_{(p,a)}(t)$ and $JC_{(p,a)}(t)$ are asymptotic (i.e., given by a divergent power series), and that there is a Stokes line: $t\in (-\infty, 0]$.  In $u$, this translates to a Stokes line along the negative real axis, $u\in (-\infty, 0]$, for the Mordell-Borel integrals in \eqref{eq:js} and \eqref{eq:jc}.
When the Borel contour grazes the Stokes line from either side, the integrals have a {\it unique}  decomposition into real and imaginary parts that corresponds precisely to the transseries decomposition. Indeed, since the formal power series of the integral is real-valued, the associated Borel function is also real for $u\in \mathbb{R}$. By the Schwarz reflection principle, the analytic continuations of this Borel function into the upper and lower half-planes are complex conjugates of each other.

Consequently, the real part of the integrals is given by the half-sum of the integrals above and below the cut, which equals the Cauchy principal value (PV) of the integrals. Since the singularities are poles, any well-behaved \'Ecalle averages coincide with the PV, thus yielding the purely perturbative part of the transseries. Therefore, the imaginary part of the integral corresponds to the nonperturbative component of the transseries.

Furthermore, the transseries decomposition is expressed in terms of unary false theta functions $\Psi_p^{(a)}(q)$ and $\Psi_p^{(a)}(\qt)$, where
\begin{eqnarray}
    q:= e^{-t} \qquad, \qquad \qt:=e^{-\pi^2/t}
    \label{eq:q-qt}.
\end{eqnarray}
The false theta functions are defined as \cite{Bring15,GMP,CCFGH,HLSS22}
\begin{equation}
\Psi^{(a)}_p (q) \;  :=  \; \sum_{n=0}^\infty \psi^{(a)}_{2p}(n) q^{\frac{n^2}{4p}} \qquad \in q^\frac{a^2}{4p}\,\mathbb{Z}[[q]]
\label{eq:false}
\end{equation}
with coefficients
\begin{eqnarray}
\psi^{(a)}_{2p}(n)  =  \left\{
\begin{array}{cl}
\pm 1, & n\equiv \pm a \quad ({\rm mod}\,\, 2p)\,, \\
0, & \text{otherwise}.
\end{array} \right.
\end{eqnarray}
We also define a related notation, extracting the fractional power of $q$,
\begin{equation}
\Psi^{(a)}_p (q) \;  :=  q^\frac{a^2}{4p}\, \Phi^{(a)}_p (q),
\label{eq:false-phi}
\end{equation}
where $\Phi^{(a)}_p (q)$ is a unary $q$-series: i.e., an expansion in positive integer powers of $q$ with coefficients $0, \pm 1$.

The unique decomposition of the Mordell-Borel integrals in \eqref{eq:js}-\eqref{eq:jc} on the Stokes line $t\in (-\infty, 0]$ is into combinations of false theta functions in terms of $1/q$ and $1/\qt$, respectively, with the following algebraic structure \cite{CDGG,ACDGO}:
\begin{eqnarray}
i \sqrt{\frac{4p|t|}{\pi}}\, JS_{(p, a)}(t) &=& q^{-\frac{a^2}{4p}}\, \Phi_{p}^{(a)}\left(\frac{1}{q}\right) +i \sqrt{\frac{\pi}{|t|}}\sum_{b=1}^{p} \sqrt{\frac{4}{p}} \sin\left(\frac{a b \pi}{p}\right)\,\qt^{\,-\frac{b^2}{p}}\Phi^{(2b)}_{2p}\left(\frac{1}{\qt^{\, 2}}\right),
\label{eq:js-false}
\end{eqnarray}
\begin{eqnarray}
i \sqrt{\frac{4p|t|}{\pi}}\, JC_{(p, a)}(t) &=&
q^{-\frac{a^2}{4p}}\,  \Phi_{2p}^{(a)}\left(\frac{1}{q^2}\right)+q^{-\frac{(2p-a)^2}{4p}}\,\Phi_{2p}^{(2p-a)}\left(\frac{1}{q^2}\right)
\nonumber\\ 
\hskip -4cm && + i\sqrt{\frac{\pi}{|t|}}\sum_{b=1}^{p} \sqrt{\frac{4}{p}} \sin\left(\frac{a (b-1/2) \pi}{p}\right)\, \qt^{\,-\frac{(2b-1)^2}{4p}}\Phi_{2p}^{(2b-1)}\left(\frac{1}{\qt^{\, 2}}\right).
\label{eq:jc-false}
\end{eqnarray}
Because of this unique decomposition when $t\in (-\infty, 0]$, into unary $q$-series (i.e., $q$-series with coefficients being $0, \pm 1$ only), we refer to the region
$q>1$ (or $t<0$) as the {\it unary side}. Correspondingly, the region with $t>0$ is referred to as the {\it non-unary side}.

On the unary side ($t<0$), the integrals, the $q$-series and $\qt$-series in \eqref{eq:js-false}-\eqref{eq:jc-false} are all resurgent functions, and the relations between them must be preserved when continued to the non-unary side ($t>0$). The resurgent preservation of relations \cite{sauzin} leads to the following conjecture \cite{CDGG,ACDGO} that for real $t>0$ there should exist unique duals $\Phi^\vee$ of the false theta functions such that
\begin{eqnarray}
 \sqrt{\frac{4pt}{\pi}}\, JS_{(p, a)}(t) &=& q^{-\frac{a^2}{4p}}\, \Phi_{p}^{(a)}\left(\frac{1}{q}\right)^\vee + \sqrt{\frac{\pi}{t}}\sum_{b=1}^{p} \sqrt{\frac{4}{p}} \sin\left(\frac{a b \pi}{p}\right)\,\qt^{\,-\frac{b^2}{p}}\Phi^{(2b)}_{2p}\left(\frac{1}{\qt^{\, 2}}\right)^\vee,
\label{eq:js-false-dual}
\end{eqnarray}
\begin{eqnarray}
\sqrt{\frac{4p t}{\pi}}\, JC_{(p, a)}(t) &=&
q^{-\frac{a^2}{4p}}\,  \Phi_{2p}^{(a)}\left(\frac{1}{q^2}\right)^\vee +q^{-\frac{(2p-a)^2}{4p}}\,\Phi_{2p}^{(2p-a)}\left(\frac{1}{q^2}\right)^\vee 
\nonumber\\ 
\hskip -4cm && + \sqrt{\frac{\pi}{t}}\sum_{b=1}^{p} \sqrt{\frac{4}{p}} \sin\left(\frac{a (b-1/2) \pi}{p}\right)\, \qt^{\,-\frac{(2b-1)^2}{4p}}\Phi_{2p}^{(2b-1)}\left(\frac{1}{\qt^{\, 2}}\right)^\vee.
\label{eq:jc-false-dual}
\end{eqnarray}
In particular, note that the algebraic structure is the same as in \eqref{eq:js-false}-\eqref{eq:jc-false}, including the mixing matrices and the exponents of the $q$ and $\qt$ powers.

We therefore identify duals to the false theta functions $\Psi^{(a)}_p(q)$ under the operation $q \to \dfrac{1}{q}$ as:
\begin{eqnarray}
\Psi^{(a)}_p(q) \longleftrightarrow  
\Psi^{(a)}_p(q)^{\vee} := q^{-\frac{a^2}{4p}}\Phi^{(a)}_{p}(q)^{\vee}
\quad,\quad   a=1, \dots, p-1.
\label{eq:false-theta-duals}
\end{eqnarray}
Note that the unique structure of the $q$-series decompositions on the unary side ($t<0$), in \eqref{eq:js-false}-\eqref{eq:jc-false}, is identical with the conjectured form on the non-unary side ($t>0$), in \eqref{eq:js-false-dual}-\eqref{eq:jc-false-dual}. However we do not know the coefficients of the dual $q$-series expansions. An important ingredient will be the knowledge of the leading non-vanishing term, which can be found numerically using the algorithm introduced in \cite{ACDGO}. The rigorous uniqueness of this procedure will be discussed in a forthcoming paper \cite{CD25}.

The numerical algorithm for generating expansions of the dual $q$-series is based on decompositions of the Mordell-Borel integrals into smaller sets that close under the modular transformation operations which probe the natural boundary:
\begin{eqnarray}
    \mathcal S\, :\, t\to \frac{\pi^2}{t}
\qquad; \qquad 
    \mathcal T\, :\, t\to t+i\, \pi.
    \label{eq:modular}
\end{eqnarray}
We refer to these smaller sets as "resurgent cyclic orbits". Two such resurgent cyclic orbits were identified in \cite{CDGG,ACDGO}, motivated by applications to Chern-Simons theory, and these are defined in the following two subsections.

\subsection{Duals of False Theta Functions}
\label{sec:false}

The full algebraic structure in \cite{ACDGO} involves 4 sets of Mordell-Borel integrals, but only the following two sets are needed to derive our analytic growth-rate estimates.
\begin{eqnarray}
J_j^{(B, p)}(t)&:=& \sqrt{2}\left(JC_{(2p, 2j-1)}(t)+(-1)^{(p-1)/2+j} JC_{(2p, 2p-(2j-1))}(t)\right),
\label{eq:b}
\\
J_j^{(D, p)}(t)&:=& JC_{(p, 2j)}(t).
\label{eq:d}
\end{eqnarray}
Each vector $J^{(*, p)}(t)$ has $\frac{(p-1)}{2}$ elements: $j=1, 2, \dots , \frac{(p-1)}{2}$. Taking linear combinations of the decompositions in \eqref{eq:js-false-dual}-\eqref{eq:jc-false-dual}, 
we obtain a decomposition into $q$ and $\qt$ series on the non-unary side:
\begin{align}
\sqrt{\ff{4pt}{\pi}}J^{(B,p)}_j(t) &= q^{-\frac{(2j-1)^2}{8p}} \Phi_{p}^{(2j-1)}(i \sqrt{q})^{\vee} + \sqrt{\ff{\pi}{t}} \sum\limits_{k=1}^{\frac{p-1}{2}} B_{jk} \, \qt^{\,-\frac{(2k-1)^2}{8p}} \Phi_{p}^{(2k-1)}(i \sqrt{\qt})^{\vee}, 
\label{eq:bq-dual}\\
\sqrt{\ff{4pt}{\pi}}J^{(D,p)}_j(t) &= q^{-
\frac{j^2}{p}} \Phi_{p}^{(2j)}(-q)^{\vee} + \sqrt{\ff{\pi}{t}} \sum\limits_{k=1}^{\frac{p-1}{2}} D_{jk} \, \qt^{\,-\frac{(2k-1)^2}{4p}} \Phi_{p}^{(2k-1)}(\qt)^{\vee}. 
\label{eq:dq-dual}
\end{align}
Note that the sums on the RHS now only extend to $k=(p-1)/2$, instead of to $b=p$, as in \eqref{eq:js-false-dual}-\eqref{eq:jc-false-dual}.
Here $\Phi^{(a)}_p(q)^{\vee}$, for $a=1, 2, ..., (p-1)$, are  $q$-series  with integer coefficients, up to an overall (rational) multiplicative factor, and the superscripts $B, D$ in \eqref{eq:b}-\eqref{eq:dq-dual} refer to the $\frac{(p-1)}{2} \times \frac{(p-1)}{2}$ mixing matrices defined as:
\begin{eqnarray}
B _{jk}&=& \frac{2}{\sqrt{p}}\sin\left( \frac{(j-1/2)(k-1/2)\pi}{p} + (-1)^{j+k+\frac{p+1}{2}} \frac{\pi}{4} \right), 
\nonumber\\
D_{jk}&=&\frac{2}{\sqrt{p}}\sin\left(\frac{j(2k-1)\pi}{p}\right).
\label{eq:reduced-mixing}
\end{eqnarray}
The important observation is that the $B$ identities in \eqref{eq:bq-dual} are vector-valued analogs of the decomposition identity in \eqref{eq:hyy}. We can therefore use the Legendre duality in Section \ref{sec:legendre}  to derive precise asymptotic estimates of the growth rate of the expansion coefficients of the dual $q$ series $\Phi^{(a)}_p(q)^{\vee}$, with odd index $a=(2j-1)$. We can then use the mixed identity \eqref{eq:dq-dual} to compute the growth rate of the expansion coefficients of the dual $q$ series $\Phi^{(2j)}_p(q)^{\vee}$, with even index $a=2j$. This therefore covers all the dual $q$-series $\Phi^{(a)}_p(q)^{\vee}$, for $a=1, \dots, (p-1)$. Comparisons with the numerical results of \cite{CDGG,ACDGO} are given below in Section \ref{sec:growth-class1}.

\subsection{Duals Associated with Brieskorn Spheres}
\label{sec:brieskorn}

Another class of vector-valued Mordell-Borel integrals (relevant for Chern-Simons theories on Brieskorn sphere manifolds), involves specially chosen linear combinations of four $JS_{(p, a)}(t)$ or $JC_{(p, a)}(t)$ integrals that form self-dual vectors under $t \rightarrow \pi^2/t$.
Two such sets were defined in \cite{ACDGO}, but we only need one set to  determine the large order growth of the related dual $q$-series. 
We define, for $p$ odd and $j=1, 2, \dots, \frac{(p-1)}{2}$:
\begin{align}
L^{(p, j)}(t)&:= \text{sign}(6j-p)JS_{(12p,|12j-2p|)}(t) + JS_{(12p,14p-12j)}(t) 
\nonumber\\
    & \qquad + JS_{(12p,12j+2p)}(t) + JS_{(12p,10p-12j)}(t). 
    \label{eq:ls}
\end{align}
We further specialize to $p=6k\pm1$, which is coprime to $2$ and $3$ for all $k$. With this choice, the integrals we define are related to the partition functions of Chern-Simons theory on the Brieskorn spheres $\Sigma(2,3,6k\pm1)$. For $k=1$ these integrals appear in the mock theta relations of mock theta functions of orders $5$ and $7$ \cite{GM12}, providing a useful test for our results. 

Taking linear combinations of the decompositions in \eqref{eq:js-false-dual}-\eqref{eq:jc-false-dual}, 
we obtain a decomposition into $q$ and $\qt$ series on the non-unary side:
\begin{eqnarray}
\sqrt{\frac{48p t}{\pi}}\, L^{(p,j)}(t) &=& q^{-2\Delta_{(p,j)}}\, X_p^{(j)}(q^2)^\vee + 
\sqrt{\frac{\pi}{t}} \sum_{k=1}^{\frac{(p-1)}{2}}  M_{jk} \, \tilde{q}^{\, -2\Delta_{(p,k)}}\, X_p^{(k)} \left(\tilde{q}^2\right)^\vee.
\label{eq:mock_q2_nonunary}
\end{eqnarray}

Here $X_p(q)^\vee$ is a $\frac{(p-1)}{2}$-component vector of $q$-series, with the $j^{th}$ entry multiplied by a rational power  $q^{-\Delta_{(p,j)}}$,  with
\begin{eqnarray}
    \Delta_{(p, j)} := \frac{(6j-p)^2}{24p}
    \qquad , \qquad j=1, 2, \dots, \frac{(p-1)}{2}.
    \label{eq:delta-mock}
    \end{eqnarray}
    The $\frac{(p-1)}{2}\times \frac{(p-1)}{2}$ mixing matrix $M$  in \eqref{eq:mock_q2_nonunary} is given by
\begin{align}
M_{jk} &= \frac{2}{\sqrt{p}}(-1)^{j+k+\lfloor \frac{p+3}{6} \rfloor}\sin{\left(\frac{6jk\pi}{p}\right)}.
\end{align}

Once again, we see that the identities in \eqref{eq:mock_q2_nonunary} constitute a vector-valued analog of the decomposition identity in \eqref{eq:hyy}. We can therefore once again use the Legendre duality in Section \ref{sec:legendre}  to derive precise asymptotic estimates of the growth rate of the expansion coefficients of the dual $q$ series $X^{(j)}_p(q)^{\vee}$. Comparisons with the numerical results of \cite{CDGG,ACDGO} are given below in Section \ref{sec:class2}.

\section{Growth Rates for Duals of False Theta Functions}
\label{sec:growth-class1}

In this section we show how the algebraic structure in \eqref{eq:bq-dual}-\eqref{eq:dq-dual},  combined with the Legendre transform argument in Section \ref{sec:legendre}, leads to analytic expressions for the growth rate of the coefficients of the $q$-series appearing in \eqref{eq:bq-dual}-\eqref{eq:dq-dual}. We present the result for general odd $p$ values, and then illustrate how it works for the order 3 ($p=3$) and order 10 ($p=5$) mock theta functions, in agreement with the known growth rates \cite{OEIS}. Then we apply the same argument to the new $q$-series for general odd $p$ and confirm that these expressions agree with the numerical results for $p=7$ and $p=9$ in \cite{ACDGO}. The data for higher $p$ values is listed in Appendix \ref{sec:results-p-class}, together with some illustrative plots.

We first note the empirical observation that the $q$-series found numerically in \cite{ACDGO} have integer coefficients which are either alternating or non-alternating in sign. This means that the most sensitive growth condition can be obtained from one of the directions approaching $q\to 1$ or $q\to -1$, whichever produces the non-alternating sign pattern.

\subsection{Growth rates for general $p$}
\label{sec:growth-p}

The algebraic structure of the decomposition identities in \eqref{eq:bq-dual}-\eqref{eq:dq-dual} almost completely fixes the asymptotic growth of the $q$-series coefficients. The only additional task is to identify the leading divergence as $t\to 0^+$ on the RHS of the decomposition identities \eqref{eq:bq-dual}-\eqref{eq:dq-dual}, which means finding the most negative power of $\qt$ on the RHS. This simply requires knowing the first term of the expansions of each of the $\Phi_p^{(a)\, \vee}$. When combined with the known exponents and mixing coefficients in the identities \eqref{eq:bq-dual}-\eqref{eq:dq-dual}, identifies the leading divergence, and using the Legendre transform result from Section \ref{sec:legendre}, this determines the large-order growth rate of the coefficients of the $q$-series expansions for all $\Phi_p^{(a)\, \vee}$.

Therefore the determination of the growth rate becomes a simple recipe, for which we use the following notation. We denote the expansions for odd and even index $a$ as:\footnote{Recall that for odd a index the expansion is in even powers of the argument, so $\Phi^{(2j-1)}_p(i\sqrt{q})^\vee$ is in fact an expansion in integer powers of $q$.}
\begin{eqnarray}
    \Phi^{(2j-1)}_p(i\sqrt{q})^\vee &=& \sum\limits_{n=\delta_{(p, 2j-1)}}^{\infty}b^{(p,2j-1)}_n q^n, 
    \label{eq:delta-pa-odd}
    \\
    \Phi^{(2j)}_p(-q)^\vee &=& \sum\limits_{n=\delta_{(p, 2j)}}^{\infty}b^{(p,2j)}_n q^n, 
    \label{eq:delta-pa-even}
\end{eqnarray}
where $\delta_{(p, 2j-1)}$ or $\delta_{(p, 2j)}$ denotes the exponent of the first nonzero term in the $q$-series, which effectively factors out leading nonzero power of $q$. Note that these exponents $\delta_{(p,a)}$ in \eqref{eq:delta-pa-odd}-\eqref{eq:delta-pa-even} are all non-negative integers.
We arrive at the following procedure for determining the leading large order growth:
\begin{enumerate}
    \item From the numerical results for the dual $q$-series, make a list of the leading exponents $\delta_{(p,a)}$, for $a=1, \dots, (p-1)$, and then compute the maximum shift:
    \begin{eqnarray}
        \tilde{\Delta}_p:=2\,{\rm max}_a\left[\Delta_{(p,a)}-\delta_{(p,a)}\right],
        \label{eq:c-recipe}
    \end{eqnarray}
    where 
    \begin{eqnarray}
        \Delta_{(p,a)}=\begin{cases}\frac{a^2}{8p}, \quad a \text{ odd} \vspace{0.5cm} \\ 
        \frac{a^2}{4p}, \quad a \text{ even}
        \end{cases}
    \end{eqnarray} 
    is the rational power accompanying the $q$-series in (\ref{eq:delta-pa-odd}) and (\ref{eq:delta-pa-even}). Then the $c$ value appearing in the large order growth of the (integer) expansion coefficients $b_n^{(p,a)}$ in \eqref{eq:delta-pa-odd}-\eqref{eq:delta-pa-even} is given by:
    \begin{eqnarray}
        c_p = \begin{cases}  \tilde{\Delta}_p/2 \quad , \quad a \text{ odd} \\ \tilde{\Delta}_p \quad , \quad a \text{ even}\end{cases}.
        \label{eq:cp}
    \end{eqnarray}
    \item 
    To determine the overall normalization factor we need also the leading coefficient in \eqref{eq:delta-pa-odd}-\eqref{eq:delta-pa-even} and the appropriate entry of the mixing matrix which multiplies the dominant growth term. Empirically, for $\Phi_p^{(a)}$ we find that the maximum in \eqref{eq:c-recipe} is achieved for an odd value of the index $a$. Denote this label as $a^*_p$. Then the corresponding  column in the mixing matrix is $j^*_p=\frac{1}{2}(a^*_p+1)$. We then define 
    \begin{eqnarray}
       \tilde{b}_p \equiv b_{\delta_{(p,a^*_p)}}^{(p,a^*_p)} 
       \label{eq:btp}
    \end{eqnarray}
    to be the leading coefficient of the $q$-series that gives the dominant growth. With this notation we find the following result for the leading growth of the $q$-series coefficients
    for the $q$-series $\Phi^{(2j-1)}_p(i \sqrt{q})^\vee$ and $\Phi^{(2j)}_p(-q)^\vee$ that appear in \eqref{eq:bq-dual}-\eqref{eq:dq-dual}:
\begin{eqnarray}
    \Phi^{(2j-1)}_p(i\sqrt{q})^\vee &=& \sum\limits_{n=\delta_{(p, 2j-1)}}^{\infty}b^{(p,2j-1)}_n q^n
    \quad \longrightarrow\quad b^{(p,2j-1)}_n \sim {\tilde{b}}_p B_{j_p j^*_p} \, \frac{e^{\pi \sqrt{2 n\,\widetilde{\Delta}_p}}}{\sqrt{4n}} ,
    \label{eq:p-growth-odd}\\
    \Phi^{(2j)}_p(-q)^\vee &=& \sum\limits_{n=\delta_{(p, 2j)}}^{\infty}b^{(p,2j)}_n q^n
    \quad \longrightarrow\quad b^{(p,2j)}_n \sim {\tilde{b}}_p \,D_{j_p j^*_p} \, \frac{e^{\pi \sqrt{4n\, \widetilde{\Delta}_p}}}{\sqrt{4n}},
    \label{eq:p-growth-even}
\end{eqnarray}
where $B_{j_p j^*_p}$ and $D_{j_p j^*_p}$ are the entries in the mixing matrix $B$ or $D$ in \eqref{eq:reduced-mixing}:
\begin{equation}
B_{j_p j_p^*}=\frac{2}{\sqrt{p}} \sin \left(\frac{\left(j_p-1 / 2\right) (j^*_p-1/2) \pi}{p}+(-1)^{j_p+j_p^*+\frac{p+1}{2}} \frac{\pi}{4}\right),
\label{eq:b*}
\end{equation}
\begin{equation}
D_{j_p j_p^*}=\frac{2}{\sqrt{p}} \sin \left(\frac{j_p (2j_p^*-1)\pi}{p}\right).
\label{eq:d*}
\end{equation}
\end{enumerate}
In the next sub-sections we present the resulting large-order growth data for $p=3, 5, 7, 9$ in Tables \ref{tab:p3}, \ref{tab:p5}, \ref{tab:p=7firstclass} and \ref{tab:p=9firstclass}, in which we list the exponents $\Delta_{(p,a)}$, the shifts $\delta_{(p, a)}$, the maximum $\tilde\Delta_p$ in \eqref{eq:c-recipe}, and the overall normalization constant $\tilde{b}_p$. In the Appendix \ref{sec:results-p-class}, we show the results for $p=11, 13, 15, 17, 19$ in Tables \ref{tab:p=11firstclass}-\ref{tab:p=19firstclass}.

We first perform the instructive exercise of verifying that the result in \eqref{eq:p-growth-odd}-\eqref{eq:p-growth-even} reproduces the known leading growth rates for the cases of $p=3$ and $p=5$, using the data in Tables \ref{tab:p3} and \ref{tab:p5}.
In the subsequent sections we apply this prescription to two new cases: $p=7$ and $p=9$. These cases are interesting because they exhibit very different rates of growth, and the reason for this can be understood in terms of the shift phenomenon described above.
 
We stress that the data required to predict the large order growth can be obtained from the known algebraic structure in \eqref{eq:bq-dual}-\eqref{eq:dq-dual}, and knowing the first term in each $q$-series at a given value of $p$, which can be found numerically using the method in \cite{ACDGO}.
In all cases these growth rate predictions agree with the large-order growth of the high-order coefficients found in \cite{ACDGO}. Sub-leading corrections are discussed in Appendix \ref{sec:sub}. These comparisons constitute a strong consistency check on the conjectured decomposition and the associated numerical algorithm proposed in \cite{ACDGO}.

\subsection{Example: $p=3$: Order 3 Mock Theta Functions}
\label{sec:p3}

The simplest example of the duals of false theta functions defined in Section \ref{sec:false} is for $p=3$. Since $\frac{(p-1)}{2}=1$, there is a scalar rather than matrix structure, and the two Mordell-Borel integrals on the LHS, defined in \eqref{eq:b}-\eqref{eq:d}, are:
{\small\begin{eqnarray}
J^{(B, 3)}(t)&=& \sqrt{2}\left(JC_{(6,1)}(t)+JC_{(6,5)}(t)\right) = \frac{ \sqrt{2}}{t}\int_0^\infty du\, e^{-6u^2/t} \frac{\cosh(5u)+\cosh(u)}{\cosh(6u)}, 
\\
J^{(D, 3)}(t)&=&JC_{(3,2)}(t) = \frac{1}{t}\int_0^\infty du\, e^{-3u^2/t} \frac{\cosh(u)}{\cosh(3u)}.
\label{eq:p3-mordell}
\end{eqnarray}}
These are the integrals $\frac{1}{\sqrt{2}} W(t)$ and $\frac{1}{2}W_2\left(\frac{t}{2}\right)$ defined on page 115 of \cite{GM12}. The decomposition identities on page 114 of \cite{GM12}, valid on the unary side $|q|<1$, can then be expressed in the form \eqref{eq:bq-dual}-\eqref{eq:dq-dual}:
\begin{eqnarray}
\sqrt{\frac{12 t}{\pi}}\,  J^{(B, 3)}(t)&=&q^{-\frac{1}{24}} \,\frac{1}{2}\, f(-q) + \sqrt{\frac{\pi}{t}} \, \qt^{\, -\frac{1}{24}} \,\frac{1}{2}\, f(-\qt),  
\label{eq:bq3-plus-growth}
\\
\sqrt{\frac{12 t}{\pi}}\,  J^{(D, 3)}(t)&=& - q^{-\frac{1}{3}} \, q\, \omega(q) +  \sqrt{\frac{\pi}{t}}    \, \qt^{\, -\frac{1}{12}} \, \frac{1}{2}\,  f(\qt^{\,2}).  
\label{eq:dq3-plus-growth}
\end{eqnarray}
This identifies the dual $q$-series in \eqref{eq:bq-dual}-\eqref{eq:dq-dual} as
\begin{eqnarray}
  \Phi_{3}^{(1)}(i\, \sqrt{q})^\vee &=& \frac{1}{2}f(-q) \sim \frac{1}{2}-\frac{q}{2}-q^2-\frac{3q^3}{2}+O(q^4),  \\
   \Phi_{3}^{(2)}(-q)^\vee &=& -q\,\omega(q)  \sim -q-2q^2-3q^3+O(q^4).
   \label{eq:p3-ids}
\end{eqnarray}
The identities \eqref{eq:bq3-plus-growth}-\eqref{eq:dq3-plus-growth} are valid for all $q$ on the non-unary side (i.e., inside the unit disk $|q|<1$). Noting that both $\frac{1}{2}f(-q)$ and $-q\,\omega(q)$ have non-alternating expansions in integer powers of $q$, this means that the fastest growth arises in the direction $q\to 1^-$; that is, $t\to 0^+$. 

In the limit $t\to 0^+$, the LHS is finite in both \eqref{eq:bq3-plus-growth} and \eqref{eq:dq3-plus-growth}:
\begin{eqnarray}
    t\to 0^+ \quad  
    \quad \Rightarrow \quad 
    \sqrt{\frac{12 t}{\pi}}\,  J^{(B, 3)}(t)\to 2
    \quad, \quad \sqrt{\frac{12 t}{\pi}}\,  J^{(D, 3)}(t)\to 1.
    \label{eq:p3-limits}
\end{eqnarray}
In this limit, $\qt\to 0^+$, and we also know that
\begin{eqnarray}
    t\to 0^+ \quad 
    \quad \Rightarrow \quad 
    \qt\to 0^+
    \quad \Rightarrow \quad 
    f(-\qt)\to 1 \quad, \quad f(\qt^{\,2})\to 1.
    \label{eq:p3-limits2}
\end{eqnarray}
Inserting these facts into \eqref{eq:bq3-plus-growth} and \eqref{eq:dq3-plus-growth} determines the leading growth of $\frac{1}{2}f(-q)$ and $-q\,\omega(q)$ in this limit:
\begin{eqnarray}
    \frac{1}{2}f(-e^{-t}) &\sim& - \frac{1}{2}\sqrt{\frac{\pi}{t}}\, e^{\frac{\pi^2}{24t}}+\dots  \qquad, \quad t\to 0^+,
    \label{eq:f-as}
    \\
    -e^{-t}\omega(e^{-t}) &\sim& - \frac{1}{2} \sqrt{\frac{\pi}{t}}\, e^{\frac{\pi^2}{12t}}+\dots  \qquad, \quad t\to 0^+.
    \label{eq:w-as}
\end{eqnarray}
We see that these both match the form of the saddle-point result \eqref{eq:gaussian}, so we can read off the parameters $b$, $c$, and $d$ in \eqref{eq:gaussian}, which therefore determines the leading growth rate of the coefficients in \eqref{eq:growth-form}. From the general results \eqref{eq:bq3-plus-growth}-\eqref{eq:dq3-plus-growth}, we find
\begin{eqnarray}
    \frac{1}{2}f(-q)=\sum_{n=0}^\infty b^{(3,1)}_n \, q^n \qquad &,& \qquad 
    b^{(3,1)}_n\sim -\frac{e^{\pi\sqrt{n/6}}}{4\sqrt{n}}\qquad, \quad n\to \infty,
    \label{eq:p3-f-an}
    \\
    -q\, \omega(q)=\sum_{n=1}^\infty b^{(3,2)}_n \, q^n \qquad &,& \qquad 
    b^{(3,2)}_n\sim -\frac{e^{\pi\sqrt{n/3}}}{4\sqrt{n}}\qquad, \quad n\to \infty,
    \label{eq:p3-w-an}
\end{eqnarray}
in agreement with the known leading growth rates in \cite{OEIS}. 

In the language of Section \ref{sec:growth-p}, the expansions of $f(q)$ and $q\, \omega(q)$ imply that the shift indices are $\delta_{(3,1)}=0$, and $\delta_{(3,2)}=1$. And since $\Delta_{(3,a)}=\frac{a^2}{12}$, we find the maximal exponent $\tilde\Delta_3=\frac{1}{12}$. Furthermore, the prefactor coefficient $\tilde{b}_p=\frac{1}{2}$. This data is shown in Table \ref{tab:p3}, and completely fixes the large order growth to be that in \eqref{eq:p3-f-an}-\eqref{eq:p3-w-an}. Thus, we see that the algebraic structure of the decomposition identities in Conjecture 1 of \cite{ACDGO}  almost completely fixes the asymptotic growth of the $q$-series coefficients. In particular, the fact that $c=\frac{1}{24}$ for $f(-q)$, and $c=\frac{1}{12}$ for $q\, \omega(q)$  in \eqref{eq:p3-f-an}-\eqref{eq:p3-w-an}, is directly related to the (negative) rational exponents of $\qt$ appearing in the identities \eqref{eq:bq3-plus-growth}-\eqref{eq:dq3-plus-growth}.
\begin{table}[h!]
\begin{tabular}{ | m{3em} | m{10em}| } 
  \hline
  $p=3$ & $\Psi_{3}^\vee$\\ 
  \hline
  $\Delta_{(3,a)}$ & $\left\{ \frac{1}{24}, \frac{1}{3}\right\}$ 
  \\
  \hline
  $\delta_{(3,a)}$ & $\left\{ 0,1 \right\}$ 
  \\
  \hline
  $\widetilde{\Delta}_3$ & $1/12$ 
  \\
  \hline
  $\tilde{b}_{3}$ & $1/2$ 
  \\
  \hline
\end{tabular}
\caption{Large order growth data for the dual $q$-series $\Psi^{(1)}_3 \left(i \sqrt{q}\right)^\vee$ and $\Psi^{(2)}_3 \left(-q\right)^\vee$. This data determines the large order growth of the coefficients in the $q$-series via the expressions in \eqref{eq:p-growth-odd}-\eqref{eq:p-growth-even}.} 
\label{tab:p3}
\end{table}
\subsection{Example: $p=5$: Order 10 Mock Theta Functions}
\label{sec:p5}
Choosing $p=5$, so that $\frac{(p-1)}{2}=2$, we have $2$-component vectors and $2\times 2$ matrix structures. The four corresponding vector-valued Mordell-Borel integrals are:
\small\begin{eqnarray}
J^{(B, 5)}(t)&=&\sqrt{2} \begin{pmatrix} JC_{(10,1)}(t)-JC_{(10,9)}(t)\\ JC_{(10,3)}(t)+JC_{(10,7)}(t)\end{pmatrix}
= 
\begin{pmatrix}
 \frac{\sqrt{2}}{t} \int_0^\infty du\, e^{-10 u^2/t} \, \frac{\cosh(9u)-\cosh(u)}{\cosh(10u)} \\
\frac{\sqrt{2}}{t} \int_0^\infty du\, e^{-10 u^2/t} \, \frac{\cosh(7u)+\cosh(3u)}{\cosh(10 u)} 
\end{pmatrix},
\label{eq:jab5}
\\
J^{(D, 5)}(t)&=& \begin{pmatrix} JC_{(5,2)}(t)\\ JC_{(5,4)}(t)\end{pmatrix} = 
\begin{pmatrix}
 \frac{1}{t} \int_0^\infty du\, e^{-5 u^2/t} \, \frac{\cosh(3u)}{\cosh(5u)} \\
\frac{1}{t} \int_0^\infty du\, e^{-5 u^2/t} \, \frac{\cosh(u)}{\cosh(5 u)} 
\end{pmatrix}.
\label{eq:jcd5}
\end{eqnarray}
Notice the different relative signs between the two terms in the $B$ class of integrals, as in \eqref{eq:b}. The integrals in \eqref{eq:jab5}-\eqref{eq:jcd5} correspond to the four integrals $J_n(t)$, ($n=4, 5, 6, 7$), appearing in the transformation identities for the order 10 mock theta functions in the notation of the Gordon-McIntosh review  (see pages 129-130 in \cite{GM12}). These identities are naturally expressed in the $2\times 2$ matrix form of \eqref{eq:bq-dual}-\eqref{eq:dq-dual}:
\begin{align}
    \sqrt{\frac{20t}{\pi}}
    J^{(B,5)}(t)
    &=
    \begin{pmatrix}
        q^{-\frac{1}{40}}X(-q)
        \\
        q^{-\frac{9}{40}}\chi(-q)
    \end{pmatrix}
    + 
   \sqrt{\frac{\pi}{t}}\frac{2}{\sqrt{5}}
    \begin{pmatrix}
        \sin(\frac{6\pi}{5}) & \sin(\frac{2\pi}{5}) \\
        \sin(\frac{2\pi}{5}) & \sin(\frac{\pi}{5})
    \end{pmatrix}
    \begin{pmatrix}
        \qt^{\, -\frac{1}{40}}\, X(-\qt) 
        \\
        \qt^{-\frac{9}{40}}\, \chi(-\qt)
    \end{pmatrix}, 
     \label{eq:jb5q}
\end{align}
\begin{align}
    \sqrt{\frac{20t}{\pi}}
    J^{(D,5)}(t)
    &=
    \begin{pmatrix}
        -q^{-\frac{1}{5}}\, \psi(q)
        \\
       - q^{-\frac{4}{5}}\, q\, \phi(q)
    \end{pmatrix}
    + 
   \sqrt{\frac{\pi}{t}}\frac{2}{\sqrt{5}}
    \begin{pmatrix}
        \sin(\frac{\pi}{5}) & \sin(\frac{3\pi}{5}) \\
        \sin(\frac{2\pi}{5}) & \sin(\frac{6\pi}{5})
    \end{pmatrix}
    \begin{pmatrix}
        \qt^{\, -\frac{1}{20}}\, X(\qt^2)
        \\
        \qt^{\, -\frac{9}{20}}\, \chi(\qt^2)
    \end{pmatrix}. 
     \label{eq:jd5q}
\end{align}
The mixing matrices correspond to the definitions in \eqref{eq:reduced-mixing}.

We therefore identify the dual $q$-series in \eqref{eq:bq-dual}-\eqref{eq:dq-dual} as:
\begin{eqnarray}
    \Phi^{(1)}_5(i\sqrt{q})^\vee =X(-q) \qquad; \qquad 
    \Phi^{(3)}_5(i\sqrt{q})^\vee =\chi(-q), \\
    \Phi^{(2)}_5(-q)^\vee =-\psi(q) \qquad ; \qquad
    \Phi^{(4)}_5(-q)^\vee =-q\, \phi(q).  
\end{eqnarray}
Note the known expansions \cite{GM12}\footnote{This information is also found by the numerical algorithm in \cite{ACDGO}.}:
\begin{eqnarray}
    X(-\qt)&=& 1+\qt+\qt^2+\dots, \\
    \chi(-\qt)&=&-\qt-\qt^2-\qt^3-2\qt^4-\dots .
\end{eqnarray}
Since $\chi(-\qt)=-\qt+\dots$, we see that on the RHS of \eqref{eq:jb5q}, the divergence as $t\to 0^+$ is coming only from the $\qt^{-\frac{1}{40}}X\left(- \,\qt \right)$ term, not the $\qt^{-\,\frac{9}{40}}\chi\left(-\qt \right)$ term. But since the integrals on the LHS are finite as $t\to 0^+$, this means that the divergences of $X(-q)$ and $\chi(-q)$ must both be cancelled by $\sqrt{\frac{\pi}{t}}\, \qt^{-\frac{1}{40}}$, but with a different trigonometric coefficient. 
Similarly, from the mixed identity \eqref{eq:jd5q} we see that the divergences of $\psi\left( q \right)$ and $\phi\left( q \right)$ are both cancelled by the $\sqrt{\frac{\pi}{t}}
\qt^{-\frac{1}{20}}X\left( \qt^2 \right)$ term, with different trigonometric coefficients. 
We therefore deduce the following results for the large-order growth of the coefficients of the dual $q$-series:
\begin{eqnarray}
    \Phi^{(1)}_5(i\sqrt{q})^\vee =X(-q) &=& \sum\limits_{n=0}^{\infty}b^{(5,1)}_n q^n, \quad b^{(5,1)}_n \sim \sin\left(\frac{\pi}{5}\right)\, \frac{e^{\pi\sqrt{n/10}}}{\sqrt{5n}}, \\
    \Phi^{(3)}_5(i\sqrt{q})^\vee =\chi(-q) &=& \sum\limits_{n=1}^{\infty}b^{(5,3)}_n q^n, \quad b^{(5,3)}_n \sim -\sin\left(\frac{2\pi}{5}\right)\, \frac{e^{\pi\sqrt{n/10}}}{\sqrt{5n}},\\
    \Phi^{(2)}_5(-q)^\vee =-\psi(q)&=& \sum\limits_{n=1}^{\infty}b^{(5,2)}_n q^n, \quad b^{(5,2)}_n \sim -\sin\left(\frac{\pi}{5}\right)\, \frac{e^{\pi\sqrt{n/5}}}{\sqrt{5n}},  \\
    \Phi^{(4)}_5(-q)^\vee =-q\, \phi(q) &=& \sum\limits_{n=1}^{\infty}b^{(5,4)}_n q^n, \quad b^{(5,4)}_n \sim -\sin\left(\frac{2\pi}{5}\right)\, \frac{e^{\pi\sqrt{n/5}}}{\sqrt{5n}}, 
\end{eqnarray}
in agreement with results in \cite{OEIS}. See also Table \ref{tab:p5}.
\begin{table}[H]
\begin{tabular}{ | m{3em} | m{10em}| } 
  \hline
  $p=5$ & $\Psi_{5}^\vee$\\ 
  \hline
  $\Delta_{(5,a)}$ & $\left\{ \frac{1}{40},\frac{1}{5},\frac{9}{40},\frac{4}{5} \right\}$ 
  \\
  \hline
  $\delta_{(5,a)}$ & $\left\{ 0,1,1,1 \right\}$ 
  \\
  \hline
  $\widetilde{\Delta}_5$ & $1/20$ 
  \\
  \hline
  $\tilde{b}_{5}$ & $1$ 
  \\
  \hline
\end{tabular}
\caption{Large order growth data for the dual $q$-series $\Psi^{(2j-1)}_5 \left(i \sqrt{q}\right)^\vee$ and $\Psi^{(2j)}_5 \left(-q\right)^\vee$. This data determines the large order growth of the coefficients in the $q$-series via the expressions in \eqref{eq:p-growth-odd}-\eqref{eq:p-growth-even}.} 
\label{tab:p5}
\end{table}
\subsection{New results: Growth Rates for Duals of $p=7$ False Thetas}
\label{sec:p7}
For $p=7$ the non-unary side decomposition identities \eqref{eq:bq-dual}-\eqref{eq:dq-dual} have a $3\times 3$ matrix structure:
\begin{eqnarray}
    \sqrt{\frac{28t}{\pi}}\, \begin{pmatrix} J^{(B,7)}_1(t) \\ J^{(B,7)}_2(t) \\ J^{(B,7)}_3(t) \end{pmatrix}
    &=& \begin{pmatrix} q^{-\frac{1}{56}} \Phi^{(1)}_7(i\sqrt{q})^{\vee} \\ 
    q^{-\frac{9}{56}} \Phi^{(3)}_7(i\sqrt{q})^{\vee} \\ 
    q^{-\frac{25}{56}} \Phi^{(5)}_7(i\sqrt{q})^{\vee} \end{pmatrix}  
    \label{eq:setsBp7mat}\\
    &+& \sqrt{\frac{\pi}{t}}\sqrt{\frac{4}{7}} 
    \begin{pmatrix}
    \sin\left(\frac{2\pi}{7}\right) & -\sin\left(\frac{\pi}{7}\right) & \sin\left(\frac{3\pi}{7}\right) \\
    -\sin\left(\frac{\pi}{7}\right) & \sin\left(\frac{3\pi}{7}\right) & \sin\left(\frac{2\pi}{7}\right) \\
    \sin\left(\frac{3\pi}{7}\right) & \sin\left(\frac{2\pi}{7}\right) & -\sin\left(\frac{\pi}{7}\right) 
    \end{pmatrix}
    \begin{pmatrix} \qt^{-\frac{1}{56}} \Phi^{(1)}_7(i\sqrt{\qt})^{\vee} \\ 
    \qt^{-\frac{9}{56}} \Phi^{(3)}_7(i\sqrt{\qt})^{\vee} \\ 
    \qt^{-\frac{25}{56}} \Phi^{(5)}_7(i\sqrt{\qt})^{\vee} \end{pmatrix},
    \nonumber\\
    \sqrt{\frac{28t}{\pi}}\, \begin{pmatrix} J^{(D,7)}_1(t) \\ J^{(D,7)}_2(t) \\ J^{(D,7)}_3(t) \end{pmatrix}
    &=& \begin{pmatrix} q^{-\frac{1}{7}} \Phi^{(2)}_7(-q)^{\vee} \\ 
    q^{-\frac{4}{7}} \Phi^{(4)}_7(-q)^{\vee} \\ 
    q^{-\frac{9}{7}} \Phi^{(6)}_7(-q)^{\vee} \end{pmatrix} 
    \nonumber\\
    &+& \sqrt{\frac{\pi}{t}} \sqrt{\frac{4}{7}}  
    \begin{pmatrix}
    \sin\left(\frac{\pi}{7}\right) & \sin\left(\frac{3\pi}{7}\right) & \sin\left(\frac{5\pi}{7}\right) \\
    \sin\left(\frac{2\pi}{7}\right) & \sin\left(\frac{6\pi}{7}\right) & \sin\left(\frac{10\pi}{7}\right) \\
    \sin\left(\frac{3\pi}{7}\right) & \sin\left(\frac{9\pi}{7}\right) & \sin\left(\frac{15\pi}{7}\right) 
    \end{pmatrix}
    \begin{pmatrix} \qt^{-\frac{1}{28}} \Phi^{(1)}_7(\qt)^{\vee} \\ 
    \qt^{-\frac{9}{28}} \Phi^{(3)}_7(\qt)^{\vee} \\ 
    \qt^{-\frac{25}{28}} \Phi^{(5)}_7(\qt)^{\vee} \end{pmatrix}.
    \label{eq:setsDp7mat}
\end{eqnarray}
The first terms of the $q$-series expansions are shown in Table \ref{tab:p7}.
\begin{center}
\begin{table}[H]
\begin{tabular}{ | m{1em} | m{7.6cm}| m{7.6cm}| } 
  \hline
  $a$ & Non-unary dual $q$-series $\Psi^{(a)}_7(q)^{\vee}$ \\ 
  \hline
  $1$ & $\frac{1}{4}q^{-1/28}\left(3+3 q^4-7 q^6+9 q^8+\dots\right)$ \\
  \hline
  $2$ & $q^{-1/7}\left(3 q - 9 q^2 + 23 q^3 - 52 q^4 +\dots\right)$ \\
  \hline
  $3$ & $\ff{1}{4}q^{-9/28}\left(1+3 q^2-6 q^4+14 q^6+\dots\right)$ \\
  \hline
  $4$ & $q^{-4/7}\left(q-3 q^2+8 q^3-17 q^4+36 q^5+\dots\right)$ \\
  \hline
  $5$ & $\ff{1}{4}q^{-25/28}\left(3 q^2-7 q^4+14 q^6-21 q^8+\dots\right)$ \\
  \hline
  $6$ & $q^{-9/7}\left(q^2-5 q^3+15 q^4-35 q^5+\dots\right)$ \\
  \hline
\end{tabular}
\caption{Small $q$ expansions of $\Psi^{(a)}_7(q)^\vee$, for $1\leq a \leq 6$.}
\label{tab:p7}
\end{table}
\end{center}
To find the growth rates of the duals of the false thetas, we look for the negative powers of $\qt$ in the second terms on the RHS. For this, we need to know the leading expansions of the duals with odd $a$ parameter. From Table \ref{tab:p7}
we have:
\begin{eqnarray}
\Phi^{(1)}_7(i \sqrt{q})^{\vee} &=& \frac{3}{4}+\frac{3}{4}q^2+\dots \quad \Rightarrow\quad \delta_{(7,1)}=0, \\
\Phi^{(3)}_7(i \sqrt{q})^{\vee} &=& \frac{1}{4}-\frac{3}{4}q^2 +\dots \quad \Rightarrow\quad \delta_{(7,3)}=0, \\
\Phi^{(5)}_7(i \sqrt{q})^{\vee} &=& -\frac{3}{4}q - \frac{7}{4}q^2 +\dots \quad \Rightarrow\quad \delta_{(7,5)}=1.
\end{eqnarray}
Therefore, the dominant negative power is $\qt^{\, -\frac{9}{28}}$ in (\ref{eq:setsDp7mat}) and $\qt^{\,-\frac{9}{56}}$ in (\ref{eq:setsBp7mat}),  not $\qt^{\, -\frac{25}{28}}$ and $\qt^{\, -\frac{25}{56}}$ because for $a=5$ the leading exponent is shifted by $+2$ and $+1$ respectively, becoming a positive power of $\qt$, which therefore vanishes, rather than diverging, in the $t\to 0^+$ limit. This data is recorded in Table \ref{tab:p=7firstclass}.
\begin{table}[H]
\begin{tabular}{ | m{3em} | m{10em}| } 
  \hline
  $p=7$ & $\Psi_{7}^\vee$ \\ 
  \hline
  $\Delta_{(7,a)}$ & $\left\{ \frac{1}{56},\frac{1}{7},\frac{9}{56},\frac{4}{7},\frac{25}{56},\frac{9}{7} \right\}$ 
  \\
  \hline
  $\delta_{(7,a)}$ & $\left\{ 0,1,0,1,1,2 \right\}$ 
  \\
  \hline
  $\widetilde{\Delta}_7$ & $9/28$ 
  \\
  \hline
  $\tilde{b}_{7}$ & $1/4$ 
  \\
  \hline
\end{tabular}
\caption{Large order growth data for the dual $q$-series $\Psi^{(2j-1)}_7 \left(i \sqrt{q}\right)^\vee$ and $\Psi^{(2j)}_7 \left(-q\right)^\vee$. This data determines the large order growth of the coefficients in the $q$-series via the expressions in \eqref{eq:p-growth-odd}-\eqref{eq:p-growth-even}.} 
\label{tab:p=7firstclass}
\end{table}
\graphicspath{ {./paper plots/} }
\begin{figure}[H]
\centering
  \includegraphics[width=0.7\linewidth]{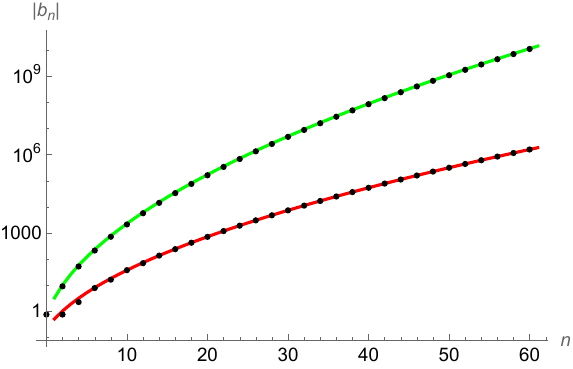}
\caption{This log plot compares the predicted growth \eqref{eq:a71growth} and \eqref{eq:a72growth}, respectively,  of the coefficients of $\Phi^{(1)}_7(i \sqrt{q})^{\vee}$ (solid red line) and $\Phi^{(2)}_7(-q)^{\vee}$ (solid green line), as in \eqref{eq:p-growth-odd}-\eqref{eq:p-growth-even}, to the coefficients determined by the numerical algorithm in \cite{ACDGO}, which are plotted as black points for both $q$-series.}
\label{fig:p_71_72growth}
\end{figure}
So in the limit $t\rightarrow0^+$ limit, the finiteness of the integrals on the LHS of \eqref{eq:setsBp7mat} implies that the $q$-series terms on the RHS diverge with the leading rate
\begin{eqnarray}
\Phi^{(2j-1)}_7(i\sqrt{q})^\vee \sim - \frac{B_{j2}}{4} \sqrt{\frac{\pi}{t}}\,\exp\left[\frac{9\pi^2}{56t}\right] \qquad, \quad t\to 0^+.
\end{eqnarray}
Here, the relevant mixing matrix entries are: $B_{j2}=\frac{2}{\sqrt{7}} \sin \left(\frac{3 \pi\left(j-\frac{1}{2}\right)}{14}+(-1)^j \frac{\pi}{4}\right)$.
Therefore, by the Legendre duality from Section \ref{sec:legendre} the
 leading large-order growth of the coefficients is:
\begin{eqnarray}
    b^{(7,1)}_n &\sim& -\frac{1}{4}\sin\left(\frac{\pi}{7}\right)\frac{e^{\pi\sqrt{9n/14}}}{\sqrt{7n}}, 
    \label{eq:a71growth}
    \\
    b^{(7,3)}_n &\sim& \frac{1}{4}\sin\left(\frac{3\pi}{7}\right) \frac{e^{\pi\sqrt{9n/14}}}{\sqrt{7n}}, \\
    b^{(7,5)}_n &\sim& \frac{1}{4}\sin\left(\frac{2\pi}{7}\right)\frac{e^{\pi\sqrt{9n/14}}}{\sqrt{7n}}.
\end{eqnarray}
A similar analysis for the decomposition identities in \eqref{eq:setsDp7mat}, whose $q$-series parts on the RHS involve the even $a$ indices, $\Phi_7^{(a)}(-q)^\vee$ with $a=2,4,6$,  gives the leading divergent behavior in the $t\rightarrow0^+$ limit as:
\begin{eqnarray}
\Phi^{(2j)}_7(-q)^\vee \sim -\frac{D_{j2}}{4} \sqrt{\frac{\pi}{t}} \exp\left[\frac{9\pi^2}{28 t} \right]
\qquad, \quad t\to 0^+.
\end{eqnarray}
Therefore the leading large-order growth of the $q$-series coefficients is
\begin{eqnarray}
    b^{(7,2j)}_n \sim \frac{1}{4}\sin\left(\frac{3j\pi}{7}\right) \frac{e^{\pi\sqrt{9n/7}}}{\sqrt{7n}}
    \qquad, \quad n\to \infty.
    \label{eq:a72growth}
\end{eqnarray}
This rapid growth is shown in Figure \ref{fig:p_71_72growth} for the $q$-series coefficients of $\Phi^{(1)}_7(i\sqrt{q})^\vee$ and $\Phi^{(2)}_7(-q)^\vee$. The expressions in \eqref{eq:a71growth} and \eqref{eq:a72growth} agree with the numerical results with high precision. Also note that the growth rates for $p=7$ are much faster than for $p=3$ or $p=5$. This fact is discussed in the next section.

\subsection{New results: Growth Rates for Duals of $p=9$ False Thetas}
\label{sec:p9}

In this Section we present the results for another new case: $p=9$. The algebraic structure of the decomposition identities \eqref{eq:bq-dual}-\eqref{eq:dq-dual} is similar to the $p=5$ and $p=7$ cases in Sections \ref{sec:p5} and \ref{sec:p7}, but now with a $4\times 4$ matrix structure. The first terms of the dual $q$-series $\Psi_9^{(a)\, \vee}$ are shown in Table \ref{tab:p9}. From these results we extract the leading exponents and the values of the leading shifted exponent $\tilde\Delta_9$, and this data is listed in Table \ref{tab:p=9firstclass}. Including the appropriate entries of the mixing matrices, we obtain analytic expressions for the leading growth rate of the coefficients of the  dual $q$-series. As seen in the general expressions \eqref{eq:p-growth-odd}-\eqref{eq:p-growth-even}, the odd $a$ and even $a$ growth rates are different. 
\begin{table}[h!]
\begin{tabular}{ | m{1em} | m{7.5cm}| m{7.5cm}| } 
  \hline
  $a$ & Non-unary dual $q$-series $\Psi^{(a)}_9(q)^{\vee}$ \\ 
  \hline
  $1$ & $q^{-1/36}\left(2+2 q^4+2 q^8-2 q^{10}+2 q^{16}+\dots\right)$ \\
  \hline
  $2$ & $q^{-1/9}\left(-2 q^2+2 q^5-2 q^6-2 q^8+2 q^9+\dots\right)$ \\
  \hline
  $3$ & $q^{-9/36}\left(2 q^2-2 q^4+2 q^{10}-2 q^{12}+\dots\right)$ \\
  \hline
  $4$ & $q^{-4/9}\left(2 q+2 q^3-2 q^4+2 q^5-2 q^6+\dots\right)$ \\
  \hline
  $5$ & $q^{-25/36}\left(-2 q^2-2 q^{10}-2 q^{18}+2 q^{20}+\dots\right)$ \\
  \hline
  $6$ & $q^{-1}\left(q-2 q^2+2 q^3-2 q^4+2 q^5-2 q^6+\dots\right)$ \\
  \hline
   $7$ & $q^{-49/36}\left(2 q^2+2 q^6-2 q^8+2 q^{10}+\dots\right)$ \\
  \hline
   $8$ & $q^{-16/9}\left(-2 q^2+2 q^3-2 q^4+2 q^{5}+\dots\right)$\\
  \hline
\end{tabular}
\caption{Small $q$ expansions of $\Psi^{(a)}_{9}(q)^\vee$, for $1\leq a \leq 8$.} 
\label{tab:p9}
\end{table}
\begin{table}[H]
\begin{tabular}{ | m{3em} | m{12em}| } 
  \hline
  $p=9$ & $\Psi_{9}^\vee$\\ 
  \hline
  $\Delta_{(9,a)}$ & $\left\{ \frac{1}{72},\frac{1}{9},\frac{9}{72},\frac{4}{9},\frac{25}{72},\frac{9}{9},\frac{49}{72},\frac{16}{9} \right\}$ 
  \\
  \hline
  $\delta_{(9,a)}$ & $\left\{ 0,2,1,1,1,1,1,2 \right\}$ 
  \\
  \hline
  $\widetilde{\Delta}_9$ & $1/36$ 
  \\
  \hline
  $\tilde{b}_{9}$ & $2$ 
  \\
  \hline
\end{tabular}
\caption{Large order growth data for the dual $q$-series $\Psi^{(2j-1)}_9 \left(i \sqrt{q}\right)^\vee$ and $\Psi^{(2j)}_9 \left(-q\right)^\vee$. This data determines the large order growth of the coefficients in the $q$-series via the expressions in \eqref{eq:p-growth-odd}-\eqref{eq:p-growth-even}.} 
\label{tab:p=9firstclass}
\end{table}
\graphicspath{ {./paper plots/} }
\begin{figure}[H]
\centering
  \includegraphics[width=0.7\linewidth]{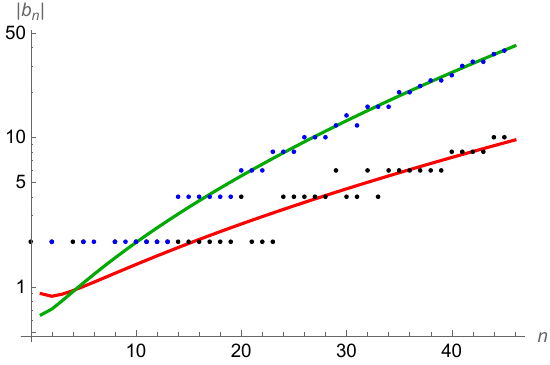}
\caption{This log plot compares the predicted growth of the coefficients of $\Phi^{(1)}_9(i \sqrt{q})$ (solid red line) and $\Phi^{(2)}_9(-q)$ (solid green line), as in \eqref{eq:p-growth-odd}-\eqref{eq:p-growth-even}, to the coefficients determined by the numerical algorithm in \cite{ACDGO}, which are plotted as black points for $\Phi^{(1)}_9(i \sqrt{q})$ and blue points for $\Phi^{(2)}_9(-q)$. Note the slow growth of the coefficients, compared to Figure \ref{fig:p_71_72growth}, and in particular the repetition of coefficient magnitudes for small $n$.}
\label{fig:p_91_92growth}
\end{figure}
Comparing the plots for $p=7$ and $p=9$, in Figures \ref{fig:p_71_72growth} and \ref{fig:p_91_92growth}, we see that the growth for $p=9$ is much slower than that for $p=7$. Furthermore, the growth for $p=7$ is much faster than for $p=5$. This sensitive dependence on $p$ is explained by our analysis. The dominant determining factor for the growth rate is the dominant negative power, $-\tilde{\Delta}_p$, of $\qt$ appearing on the RHS of the decomposition identities \eqref{eq:bq-dual}-\eqref{eq:dq-dual}. This, in turn is determined by the difference between $\frac{a^2}{4p}$ and the exponent of the first non-zero term in the $q$-series. In Figure \ref{fig:histogram1} we show a histogram of the values of  $\tilde{\Delta}_p$, as a function of (odd) $p$ values. We see that these values fluctuate significantly. And we observe that for $p=7$ the value is much larger than for $p=5$ or for $p=9$. This is the reason for the dramatically different rates of growth. This pattern explains the different rates of growth of the numerically determined coefficients found in \cite{ACDGO}, as summarized in Tables in Appendix \ref{sec:results-p-class}.
It would be interesting to understand in more detail the physical origin of these shift indices, given that $\tilde{\Delta}_p$ is related to an effective central charge in a three-dimensional $\mathcal N=2$ supersymmetric QFT \cite{GJ}.

Another interesting feature of the results for $p=9$ values can be seen in Figure \ref{fig:p_91_92growth}, where many of the low order coefficients have the same magnitude. There are many repetitions of coefficients of magnitude $2$, $4$, $6$, etc, before the pattern settles down to the asymptotic value. This "degeneracy of degeneracies" is a feature of the very slowly growing coefficients, since only by repetition can an integer sequence grow slowly. Small, slowly growing integer coefficients cannot match a smooth curve: this is at the core of the observed low-order scatter in this and other examples discussed below (see Figures \ref{fig:mock5-growth}, \ref{fig:mock7-growth} and \ref{fig:mock13-growth}). In the next section, we compare our results for the duals of false theta functions with other results in the literature, and discuss the connection between this slow growth and the condition of ''optimality" in the $q$-series literature.
\begin{figure}[H]
\centering
  \includegraphics[width=0.65\linewidth]{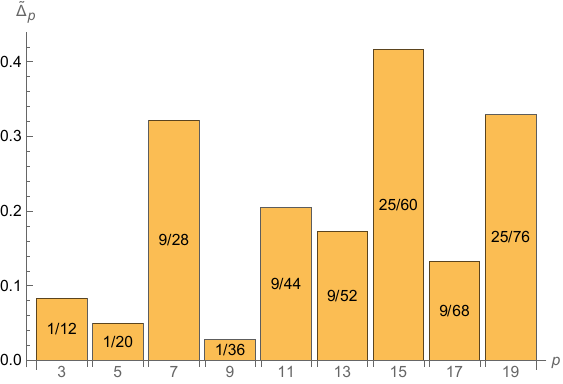}
\caption{Values of $\tilde{\Delta}_p$, the key parameter defined in \eqref{eq:c-recipe} for the duals of false theta functions, for odd $p$ values. This determines $c_{\rm eff}$ as in \eqref{eq:cp}. Note the wide variation of values. Smaller $\tilde{\Delta}_p$ corresponds to slower growth of the $q$-series coefficients.}
  \label{fig:histogram1}
\end{figure}
\subsection{Comparisons with other results for Duals of False Theta Functions}
In this section we compare our new results for duals of false theta functions with other constructions in the $q$-series literature. As is discussed in the previous section, for $p=3$ and $p=5$ we find growth rates consistent with results given in \cite{OEIS}. For most of our new results $(p>5)$ on duals of false thetas, we do not find any $q$-series in the literature that match ours, with the exception of the $p=9$ duals in Section \ref{sec:p9}. These $p=9$ $q$-series, along with those found for $p=3$ and $p=5$, have \textit{optimal} growth, first introduced in \cite{DMZ12} and formalized by Cheng and Duncan in \cite{CD20c}, with further discussion in \cite{CDH13, CCFGH}. The optimality condition in \cite{DMZ12} states that the $q$-series have coefficients that grow \textit{as slowly as possible}, and we see from our cancellation argument that the slowest possible growth corresponds exactly to $\tilde{\Delta}_p=\frac{1}{4p}$. See Figure \ref{fig:histogram1}. This is equivalent to the statement in \cite{CD20c} that optimality is the condition
\begin{eqnarray}
    Y(q) \sim O\left(q^{-\frac{1}{4p}}\right), \quad t \to 0
\end{eqnarray}
Furthermore, in these $p=3$, $p=5$ and $p=9$ cases, the $q$-series coefficients match some of the coefficients $C^{(m+K)}(r^2-4mn,r)$ given in Tables 3 - 18 in \cite{CD20c}, where $r$ is the indexing of the mock Jacobi theta functions. In particular,
\begin{itemize}
    \item The coefficients of $-4\Phi^{(1)}_3(\sqrt{q})^\vee$ match the coefficients $C^{(6+2)}(r^2-24n,r)$ for $r=1$, given in Table $3$ of \cite{CD20c}. There seem to be no tables in \cite{CD20c} with coefficients related to $\Phi^{(2)}_3(q)^\vee$.
    \item The coefficients of $\Phi^{(a)\,\vee}_5$ for $a=2, 4$ match the coefficients $C^{(20+4)}(r^2-80n,r)$, up to an overall factor, for $r=4, 8$, given in Table $9$ of \cite{CD20c}, and for $a=1,3$ they match $C^{(10+2)}(r^2-40n,r)$ up to an overall factor for $r=1,3$, given in Table $4$ of \cite{CD20c}.
    \item The coefficients $C^{(18+2)}(r^2-72n,r)$ in Table 6 of \cite{CD20c} for $r=1,3,5,7$ match those of $-\Phi^{(a)}_9(\sqrt{q})^\vee$ for $a=1,3,5,7$ respectively. We also find that the coefficients $C^{(36+4)}(r^2-144n,r)$ in Table 14 in \cite{CD20c} for $r=4,8,12,16$ match those of $-\Phi^{(a)}_9(-q)^\vee$ for $a=2,4,6,8$ respectively.
\end{itemize}
To emphasize the matching with the optimal mock Jacobi function coefficients for $p=9$, in Table \ref{tab:p9optimal}  we present the $q$-series expansions of $-\Phi^{(a)}_9(\sqrt{q})^\vee$ for odd $a$, and $-\Phi^{(a)}(-q)^\vee$ for even $a$, as well as the coefficients $C^{(18+2)}(r^2-72n,r)$ from \cite{CD20c}, for $r=1,3,5,7$ and $C^{(36+4)}(r^2-144n,r)$ for $r=4,8,12,16$.
\begin{table}[h!]
\begin{tabular}{ | m{1em} | m{7cm}| m{1em} | m{7.5cm}| } 
  \hline
  $a$ & $-\Phi^{(a)}_9(\sqrt{q})$ & $a$ & $-\Phi^{(a)}_9(-q)^{\vee}$ \\ 
  \hline
  $1$ & $-2-2 q^2-2 q^4+2 q^{5}-2 q^{8}+\dots$ & $2$ & $2 q^2+2 q^5+2 q^6+2 q^8+2 q^9+\dots$ \\
  \hline
  $3$ & $-2 q + 2 q^2 - 2 q^5 + 2 q^6 - 2 q^7 + 2 q^8+\dots$ & $4$ & $2 q + 2 q^3 + 2 q^4 + 2 q^5 + 2 q^6 + 2 q^7 + 2 q^8+\dots$ \\
  \hline
  $5$ & $2 q+2 q^{5}+2 q^{9}-2 q^{10}+\dots$ & $6$ & $q + 2 q^2 + 2 q^3 + 2 q^4 + 2 q^5 + 2 q^6 + 4 q^7 + 4 q^8+\dots$ \\
  \hline
  $7$ & $-2 q - 2 q^3 + 2 q^4 - 2 q^5 - 2 q^7 + 2 q^8+\dots$ & $8$ & $2 q^2 + 2 q^3 + 2 q^4 + 2 q^5 + 4 q^6 + 2 q^7 + 4 q^8+\dots$ \\
  \hline
\end{tabular}
\\ \vspace{0.5cm}
\begin{tabular}{ | m{1.5em} | m{1em} m{1em} m{1em} m{1em} m{1em} m{1em} m{1em} m{1em} m{1.5em} | m{1.5em} |  m{1em} m{1em} m{1em} m{1em} m{1em} m{1em} m{1em} m{1em} m{1.5em} | } 
  \hline
  $r,n$ & 0 & 1 & 2 & 3 & 4 & 5 & 6 & 7 & 8 & $r,n$ & 0 & 1 & 2 & 3 & 4 & 5 & 6 & 7 & 8  \\ 
  \hline
  $1$ & -2 & 0 & -2 & 0 & -2 & 2 & 0 & 0 & -2 & 4 & 0 & 0 & 2 & 0 & 0 & 2 & 2 & 0 & 2 \\
  \hline
  $3$ & 0 & -2 & 2 & 0 & 0 & -2 & 2 & -2 & 2 & 8 & 0 & 2 & 0 & 2 & 2 & 2 & 2 & 2 & 2 \\
  \hline
  $5$ & 0 & 2 & 0 & 0 & 0 & 2 & 0 & 0 & 0 & 12 & 0 & 1 & 2 & 2 & 2 & 2 & 2 & 4 & 4\\
  \hline
  $7$ & 0 & -2 & 0 & -2 & 2 & -2 & 0 & -2 & 2 & 16 & 0 & 0 & 2 & 2 & 2 & 2 & 4 & 2 & 4 \\
  \hline
\end{tabular}
\caption{Top: Rescaled $q$-series expansions for $p=9$, where the left side lists the odd $a$ series, and the right side lists the even $a$ series as obtained by numerical methods in \cite{ACDGO}. Bottom: Fourier coefficients $C^{(18+2)}(r^2-72n,r)$ for $r=1,3,5,7$ and $C^{(36+4)}(r^2-144n,r)$ for $r=4,8,12,16$ up to $n=8$ from Tables 6 and 14 in \cite{CD20c} respectively.} 
\label{tab:p9optimal}
\end{table}
\section{Growth Rates for Duals for Brieskorn Spheres}
\label{sec:class2}
The second resurgent cyclic orbit of Mordell-Borel integrals that we study in this paper generalize those appearing in the order $5$ and order $7$ mock theta identities in \cite{GM12}. We expect that the $q$-series obtained from decomposing these integrals are related to the quantum $\widehat{Z}$ invariants of the orientation-reversed Brieskorn spheres $\overline{\Sigma(2,3,6k\pm1)}$. We will label these examples with an integer $p = 6k\pm1$. While the case of $k=1$ $(p=5,7)$ has been explored extensively in the literature, results from \cite{ACDGO} allow us to study these $q$-series for larger $k$.

In \cite{ACDGO}, the approach of imposing preservation of relations when passing from the unary to the non-unary side was applied to these orientation-reversed Brieskorn spheres 
$\overline{\Sigma(2,3,6k\pm1)}$. Then, as for the duals of the false theta functions analyzed in 
section \ref{sec:growth-class1}, the algebraic structure (and knowledge of the leading power of each $q$-series), results in the following prediction for the growth rates of the $q$-series coefficients:
\begin{eqnarray}
    X^{(j)}_{(2,3,p)}(q)^\vee = \sum\limits_{n=\delta_{(p,j)}}^{\infty} b_n^{(j)} q^n \qquad \longrightarrow \qquad b^{(j)}_n \sim \tilde{b}_{(2,3,p)} M_{jj^*} \frac{e^{\pi \sqrt{16\widetilde{\Delta}_{(2,3,p)}\, n}}}{\sqrt{2n}},
    \label{eq:mockgrowth}
\end{eqnarray}
where $\widetilde{\Delta}_{p}$, $\tilde{b}_p$, and $j^*$ are defined analogously to the previous examples. By comparing these asymptotics to the discussion in \cite{GJ}, we see that the contribution of the half-index to the effective central charge of the $3d$ superconformal index (\ref{eq:index}) from this growth is given by
\begin{eqnarray}
c_{\rm eff,\frac{1}{2}}^{T[\overline{\Sigma(p_1,p_2,p_3)}]}=1+24\tilde{\Delta}_{(p_1,p_2,p_3)}.
\end{eqnarray}
We then recover the full effective central charge of the $3d$ index by including the trivial contribution from $T[\Sigma(p_1,p_2,p_3)]$, giving
\begin{eqnarray}
c_{\rm eff}^{T[{\Sigma(p_1,p_2,p_3)}]}=2+24\tilde{\Delta}_{(p_1,p_2,p_3)}.
\end{eqnarray}
Note that we generalize this notation to any orientation-reversed Brieskorn sphere $\overline{\Sigma(p_1,p_2,p_3)}$.
In this Section we first confirm the behavior of the growth for the known examples of $\overline{\Sigma(2,3,5)}$ and $\overline{\Sigma(2,3,7)}$, for which the dual $q$-series are in terms of mock theta functions of order 5 and order 7, respectively. Then we analyze the new results from \cite{ACDGO}, for $\overline{\Sigma(2,3,11)}$ and $\overline{\Sigma(2,3,13)}$. Results for $\overline{\Sigma(2,3,17)}$ and $\overline{\Sigma(2,3,19)}$ are listed in Appendix \ref{sec:results-mock-class}. After reviewing these cases in depth, we give a larger list of results and compare the effective central charges with other conjectured results in the literature. While much of the discussion is focused on the class $\overline{\Sigma(2,3,6k\pm1)}$, we can also use methods developed in \cite{ACDGO} to extract central charge data for any Brieskorn sphere. For these examples we will only discuss the asymptotic data $\tilde{\Delta}_{(p_1,p_2,p_3)}$, as the form of the mixing matrices appearing in the asymptotic growth is not known in general and not relevant to finding the effective central charges.

\subsection{Example: $\overline{\Sigma (2,3,5)}$ and Order 5 Mock Theta Functions}
\label{sec:mock5}

For $p=5$, the integral vectors defined in \eqref{eq:ls} have 2 components, 
\begin{eqnarray}
\begin{pmatrix}
    L^{(5,1)}(t)
    \\
    L^{(5,2)}(t)
\end{pmatrix}
=  \begin{pmatrix}
    \frac{1}{t}\int_0^\infty du\, e^{-60 u^2/t} \left[\frac{\sinh(2u)+\sinh(22 u)+\sinh(38 u)+\sinh(58 u)}{\sinh(60 u)}\right]
    \\
    \frac{1}{t}\int_0^\infty du\, e^{-60 u^2/t} \left[\frac{\sinh(14 u)+\sinh(26 u)+\sinh(34 u)+\sinh(46 u)}{\sinh(60 u)}\right]
\end{pmatrix}.
    \label{eq:mock5-integrals}
\end{eqnarray}
The corresponding decomposition identities in \cite{GM12} for order 5 mock theta functions are:
\begin{eqnarray}
\hskip -1cm 
\sqrt{\frac{540t}{\pi}} 
\begin{pmatrix}
L^{(5,1)}(t)
\\
L^{(5,2)}(t)
\end{pmatrix}
=
\begin{pmatrix}
-q^{-\frac{1}{60}} (\chi_0(q^2)-2)
\\
-q^{-\frac{49}{60}} q^2\, \chi_1(q^2)
\end{pmatrix}
+\sqrt{\frac{\pi}{t}}  \frac{2}{\sqrt{5}} 
\begin{pmatrix}
\sin\left(\frac{\pi}{5}\right) & \sin\left(\frac{2\pi}{5}\right)
\\
\sin\left(\frac{2\pi}{5}\right) & - \sin\left(\frac{\pi}{5}\right) 
\end{pmatrix}
\begin{pmatrix}
-\tilde{q}^{\, -\frac{1}{60}} (\chi_0(\tilde{q}^2)-2)
\\
-\tilde{q}^{\, -\frac{49}{60}} \tilde{q}^2 \, \chi_1(\tilde{q}^2)
\end{pmatrix}.
\label{eq:mock5-chi02-chi12}
\end{eqnarray}
\graphicspath{ {./paper plots/} }
\begin{figure}[H]
\centering
  \includegraphics[width=0.7\linewidth]{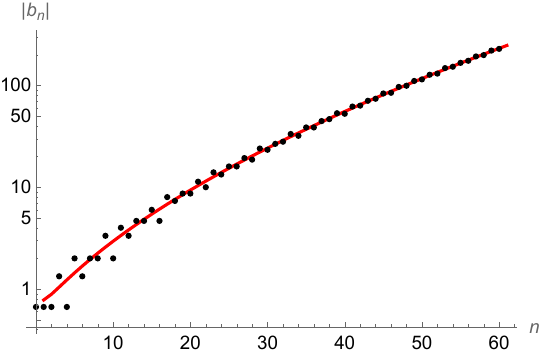}
\caption{This log plot compares the predicted growth [solid red line]
for the $q$-series coefficients of $X^{(1)}_{(2,3,5)}(q)^\vee$, as in \eqref{eq:mockgrowth}-\eqref{eq:mockgrowth5}, with the coefficients determined by the numerical algorithm in \cite{ACDGO}, which are plotted as black points.
}
\label{fig:mock5-growth}
\end{figure}
Comparing with the decomposition identities in \eqref{eq:mock_q2_nonunary} leads to the identifications 
\begin{eqnarray}
    X^{(1)}_{(2,3,5)}(q)^\vee &=& -\frac{2}{3}\left(\chi_0(q)-2\right) 
    = -\frac{2}{3}\left( -1+q+q^2+2q^3+q^4+\dots\right),
    \\
    X^{(2)}_{(2,3,5)}(q)^\vee &=& -\frac{2}{3}q\,\chi_1(q) 
    = -\frac{2}{3}\left( q+2q^2+2q^3+3q^4+3q^5+
    \dots\right).
\end{eqnarray}
\begin{table}[H]
\begin{tabular}{ | m{3em} | m{10em}| } 
  \hline
  $p=5$  & $X_{(2,3,5)}(q)^\vee$\\ 
  \hline
  $\Delta_{(5,a)}$ & $\left\{ \frac{1}{120},\frac{49}{120}\right\}$ 
  \\
  \hline
  $\delta_{(5,a)}$ & $\left\{ 0,1 \right\}$ 
  \\
  \hline
  $\tilde{b}_{(2,3,5)}$ & $-2/3$
  \\
  \hline
  $\widetilde{\Delta}_{(2,3,5)}$ & $1/120$ 
  \\
  \hline
  $c_{\rm eff,\frac{1}{2}}^{T[\overline{M_3}]}$ & $6/5$
  \\
  \hline
\end{tabular}
\caption{Large order growth data, for expression \eqref{eq:mockgrowth}, for the $q$-series $X^{(1)}_{(2,3,5)}(q)^\vee = -\frac{2}{3}\left(\chi_0(q)-2\right)$ and $X^{(2)}_{(2,3,5)}(q)^\vee = -\frac{2}{3}q\,\chi_1(q)$.} 
\label{tab:mock5}
\end{table}
Therefore, recalling that the $q$-series in \eqref{eq:mock_q2_nonunary} have argument $q^2$ and $\qt^2$, we see that the index shifts are $\delta_{(5,1)}=0$ and $\delta_{(5,2)}=2$ for $X^{(1)}_{(2,3,5)}(q^2)^\vee$ and $X^{(1)}_{(2,3,5)}(q^2)^\vee$, respectively. Therefore the dominant negative power is $\qt^{-1/120}$, as indicated in Table \ref{tab:mock5}. Hence the growth rates are:
\begin{eqnarray}
    X^{(j)}_{(2,3,5)}(q)^\vee = \sum\limits_{n=\delta_{(5,j)}}^{\infty} b_n^{(5,j)} q^n 
    \qquad \longrightarrow \qquad 
    b^{(5,j)}_n \sim -\frac{2}{3} \sin\left(\frac{j \pi}{5}\right) \sqrt{\frac{4}{5}}\frac{e^{\pi \sqrt{16\, n/120}}}{\sqrt{2n}}.
    \label{eq:mockgrowth5}
\end{eqnarray}
See Figure \ref{fig:mock5-growth}. Note the scatter of the coefficients at low orders, before settling down to the leading asymptotic growth rate in \eqref{eq:mockgrowth5}.

\subsection{Example: $\overline{\Sigma (2,3,7)}$ and Order 7 Mock Theta Functions}
\label{sec:mock7}

\begin{figure}[h!]
\centering
  \includegraphics[width=0.7\linewidth]{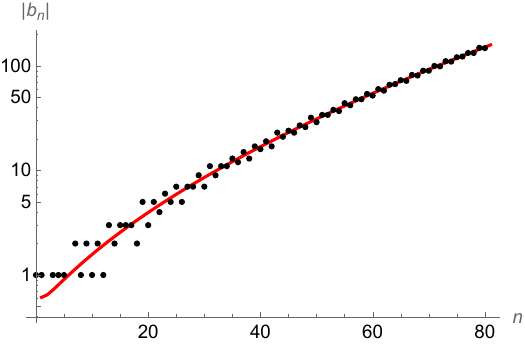}
\caption{This log plot compares the predicted growth [solid red line]
for the $q$-series coefficients of $X^{(1)}_{(2,3,7)}(q)^\vee$, as in \eqref{eq:mockgrowth}-\eqref{eq:mockgrowth7}, with the coefficients determined by the numerical algorithm in \cite{ACDGO}, which are plotted as black points.
}
\label{fig:mock7-growth}
\end{figure}

For $p=7$, the integral vectors defined in \eqref{eq:ls} have $\frac{(p-1)}{2}=3$ components: 
\begin{equation}
\begin{pmatrix}
    L^{(7,1)}(t)
    \\
    L^{(7,2)}(t)
    \\
    L^{(7,3)}(t)
\end{pmatrix}
=\begin{pmatrix}
    \frac{1}{t}\int_0^\infty du\, e^{-84 \frac{u^2}{t}}\left[\frac{-\sinh(2u)+\sinh(26 u)+\sinh(58 u)-\sinh(82 u)}{\sinh(84 u)}\right]
    \\
    \frac{1}{t}\int_0^\infty du\, e^{-84 \frac{u^2}{t}} \left[\frac{\sinh(10 u)+\sinh(38 u)+\sinh(46 u)+\sinh(74 u)}{\sinh(84 u)}\right]
    \\
    \frac{1}{t}\int_0^\infty du\, e^{-84 \frac{u^2}{t}} \left[\frac{\sinh(22 u)+\sinh(34 u)+\sinh(50 u)+\sinh(62 u)}{\sinh(84 u)}\right]
\end{pmatrix}.
\label{eq:mock7-integralsL}
\end{equation}
The corresponding decomposition identities in \cite{GM12} for order 7 mock theta functions can be written as:
\begin{eqnarray}
\hskip -1cm 
\sqrt{\frac{336 t}{\pi}} 
\begin{pmatrix}
L^{(7,1)}(t)
\\
L^{(7,2)}(t)
\\
L^{(7,3)}(t)
\end{pmatrix}
&=&
\begin{pmatrix}
-q^{-\frac{1}{84}}\mathcal F_0(q^2)
\\
-q^{-\frac{25}{84}}\mathcal F_1(q^2)
\\
-q^{-\frac{121}{84}} q^2 \mathcal F_1(q^2)
\end{pmatrix} 
\nonumber\\
&\hskip -4cm + & \hskip -2cm
\sqrt{\frac{\pi}{t}}  \frac{2}{\sqrt{7}} 
\begin{pmatrix}
\sin\left(\frac{\pi}{7}\right) & \sin\left(\frac{2\pi}{7}\right) & \sin\left(\frac{3\pi}{7}\right)
\\
\sin\left(\frac{2\pi}{7}\right) &  -\sin\left(\frac{3\pi}{7}\right) & \sin\left(\frac{\pi}{7}\right)
\\
\sin\left(\frac{3\pi}{7}\right) &  \sin\left(\frac{\pi}{7}\right) & -\sin\left(\frac{2\pi}{7}\right)
\end{pmatrix}
\begin{pmatrix}
-\qt^{\, -\frac{1}{84}}\mathcal F_0(\qt^2)
\\
-\qt^{\, -\frac{25}{84}}\mathcal F_1(\qt^2)
\\
-\qt^{\, -\frac{121}{84}} \qt^2 \mathcal F_1(\qt^2)
\end{pmatrix}.
\label{eq:l-7}
\end{eqnarray}
Comparing with the decomposition identities in \eqref{eq:mock_q2_nonunary} leads to the identifications 
\begin{eqnarray}
    X^{(1)}_{(2,3,7)}(q)^\vee &=& -\mathcal{F}_0(q)
    = -1-q-q^3-q^4-q^5-2q^7 + \dots,
    \\
    X^{(2)}_{(2,3,7)}(q)^\vee &=& -\mathcal{F}_1(q) 
     =-q-q^2 - q^3 - q^4 - 2 q^5 - q^6 - 2 q^7 + \dots,
     \\
    X^{(3)}_{(2,3,7)}(q)^\vee &=& -q\, \mathcal{F}_2(q) 
     =-q - q^2 - 2 q^3 - q^4 - 2 q^5 - 2 q^6 + \dots.
\end{eqnarray}
Therefore the dominant negative power is $\qt^{-1/168}$, as indicated in Table \ref{tab:mock7}. Hence the growth rates are:
\begin{eqnarray}
    X^{(j)}_{(2,3,7)}(q)^\vee = \sum\limits_{n=\delta_{(7,j)}}^{\infty} b_n^{(7,j)} q^n
    \qquad \longrightarrow \qquad
    b^{(7,j)}_n \sim  -\sin\left(\frac{j \pi}{7}\right) \sqrt{\frac{4}{7}}  \frac{e^{\pi \sqrt{16\, n/168}}}{\sqrt{2n}}.
    \label{eq:mockgrowth7}
\end{eqnarray}
These growth rates agree with the known growth rates of the coefficients of the order 7 mock theta functions \cite{OEIS}.
See Figure \ref{fig:mock7-growth}.  The behavior is similar to that found for $p=5$, as shown in Figure \ref{fig:mock5-growth}. Observe once again the scattered behavior at low order, which settles down to the leading asymptotic growth rate at large order.
\begin{table}[h!]
\begin{tabular}{ | m{3em} | m{10em}| } 
  \hline
  $p=7$  & $X_{(2,3,7)}(q)^\vee$ \\ 
  \hline
  $\Delta_{(7,a)}$ & $\left\{ \frac{1}{168},\frac{25}{168},\frac{121}{168} \right\}$ 
  \\
  \hline
  $\delta_{(7,a)}$ & $\left\{ 0,1,1 \right\}$ 
  \\
  \hline
  $\tilde{b}_{(2,3,7)}$ & $-1$ 
  \\
  \hline
  $\widetilde{\Delta}_{(2,3,7)}$ & $1/168$   
  \\
  \hline
  $c_{\rm eff, \frac{1}{2}}^{T[\overline{M_3}]}$ & $8/7$
  \\
  \hline
\end{tabular}
\caption{Large order growth data, for the growth rate expression \eqref{eq:mockgrowth}, for the $q$-series $X^{(1)}_{(2,3,7)}(q)^\vee = -\mathcal{F}_0(q)$, $X^{(2)}_{(2,3,7)}(q)^\vee = -\mathcal{F}_1(q)$, and $X^{(3)}_{(2,3,7)}(q)^\vee = -q\,\mathcal{F}_2(q)$.} 
\label{tab:mock7}
\end{table}
\graphicspath{ {./paper plots/} }
\subsection{New Results: Growth Rates for $\overline{\Sigma(2,3,11)}$}
\label{sec:mock11}
\begin{figure}[H]
\centering
  \includegraphics[width=0.7\linewidth]{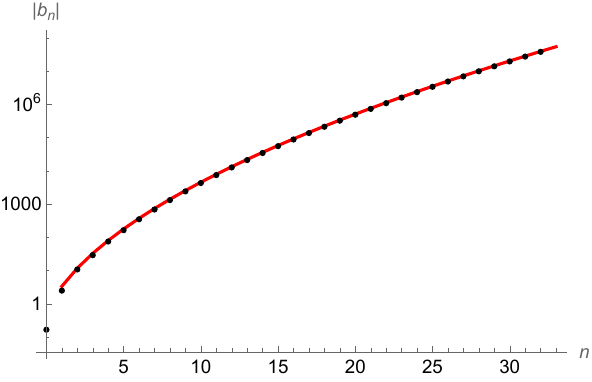}
\caption{This log plot compares the predicted growth [solid red line]
for the $q$-series coefficients of $X^{(1)}_{(2,3,11)}(q)^\vee$, as in \eqref{eq:mockgrowth}, with the coefficients determined by the numerical algorithm in \cite{ACDGO}, which are plotted as black points.}
\label{fig:mock11-growth}
\end{figure}

For $p=11$, the integral vectors defined in \eqref{eq:ls} have $\frac{(p-1)}{2}=5$ components. Unlike the $p=5$ and $p=7$ cases, for $p=11$ there are no previously known Mordell-Borel decomposition identities with which to compare. However, in \cite{ACDGO} the $p=11$ decomposition in \eqref{eq:mock_q2_nonunary} was solved numerically, leading to dual $q$-series, the first few terms of which are listed in Table \ref{tab:mock11-table-match}. In \cite{ACDGO} many more terms of these $q$-series are presented. 
\begin{table}[H]
\centering
\begin{tabular}{ | m{1em} | m{8.2cm}| } 
  \hline
  $j$ & Non-unary dual $q$-series $X^{(j)}_{(2,3,11)}(q)^{\vee}$  \\ 
  \hline
  $1$ & $\frac{1}{6}q^{-25/264}\left(-1+15 q+65 q^2+175 q^3+\dots\right)$ 
  \\
  \hline
  $2$ & $\frac{1}{6}q^{-1/264}\left(5+5 q+15 q^2+60 q^3+125 q^4+\dots\right)$ 
  \\
  \hline
  $3$ & $\frac{1}{6}q^{-49/264}\left(-16 q-55 q^2-155 q^3-385 q^4+\dots\right)$ 
  \\
  \hline
  $4$ & $\frac{1}{6}q^{-169/264}\left(-5 q-22 q^2-60 q^3-155 q^4+\dots\right)$ 
  \\
  \hline
  $5$ & $\frac{1}{6}q^{-361/264}\left(5 q^2+33 q^3+99 q^4+268 q^5+\dots\right)$ 
  \\
  \hline
\end{tabular}
\caption{Small $q$ expansions of the $p=11$ dual $q$-series $X^{(j)}_{(2,3,11)}(q)^\vee$, for $1\leq j \leq 5$, for the Brieskorn sphere $\overline{\Sigma(2, 3, 11)}$.} 
\label{tab:mock11-table-match}
\end{table}
Using our general expression \eqref{eq:mockgrowth}, the growth rate data is shown in Table \ref{tab:p=11secondclass}. The comparison with the actual coefficients is shown in Figure \ref{fig:mock11-growth} for $(p,j)=(11,1)$, showing excellent agreement. Similar results apply for $j=2,3, \dots ,(p-1)$, although subleading corrections become more pronounced for higher $j$ values; see Section \ref{sec:sub} below. Also note that the growth is much faster for $p=11$ than for $p=5$ or $p=7$, as can be seen comparing Figure \ref{fig:mock11-growth} with Figure \ref{fig:mock5-growth} and Figure \ref{fig:mock7-growth}. Correspondingly, there is much less scatter in the coefficients at low order. This is discussed further in the next section.
\begin{table}[H]
\begin{tabular}{ | m{3em} | m{10em}| } 
  \hline
  $p=11$  & $X_{(2,3,11)}(q)^\vee$ \\ 
  \hline
  $\Delta_{(11,a)}$ & $\left\{ \frac{25}{264},\frac{1}{264},\frac{49}{264},\frac{169}{264},\frac{361}{264} \right\}$ 
  \\
  \hline
  $\delta_{(11,a)}$ & $\left\{ 0,0,1,1,2 \right\}$ 
  \\
  \hline
  $\tilde{b}_{(2,3,11)}$ & $1/6$  
  \\
  \hline
  $\widetilde{\Delta}_{(2,3,11)}$ & $25/264$
  \\
  \hline
  $c_{\rm eff,\frac{1}{2}}^{T[\overline{M_3}]}$ & $36/11$
  \\
  \hline
\end{tabular}
\caption{Large order growth data, for the growth rate expression \eqref{eq:mockgrowth}, for the dual $q$-series $X^{(j)}_{(2,3,11)}(q)^\vee$, with $j=1, 2, \dots, 5$.} 
\label{tab:p=11secondclass}
\end{table}
\graphicspath{ {./paper plots/} }
\subsection{New Results: Growth Rates for $\overline{\Sigma(2,3,13)}$}
\label{sec:mock13}
For $p=13$, the integral vectors defined in \eqref{eq:ls} have $\frac{(p-1)}{2}=6$ components. Once again, unlike the $p=5$ and $p=7$ cases, for $p=13$ there are no previously known Mordell-Borel decomposition identities with which to compare. In \cite{ACDGO} the $p=13$ decomposition in \eqref{eq:mock_q2_nonunary} was solved numerically, leading to dual $q$-series, the first few terms of which are listed in Table \ref{tab:mock13-table-match}. In \cite{ACDGO} many more terms of these $q$-series are presented. 
\begin{table}[H]
\begin{center}
\begin{tabular}{ | m{1em} | m{7cm}| } 
  \hline
  $j$ & Non-unary dual $q$-series $X^{(j)}_{(2,3,13)}(q)^{\vee}$  \\ 
  \hline
  $1$ & $q^{-49/312}\left(-2 q-2 q^7-2 q^9-2 q^{13}+\dots\right)$ 
  \\
  \hline
  $2$ & $q^{-1/312}\left(-2-2 q^3-2 q^4-2 q^7+\dots\right)$ 
  \\
  \hline
  $3$ & $q^{-25/312}\left(-2 q-2 q^3-2 q^5-2 q^7+\dots\right)$ 
  \\
  \hline
  $4$ & $q^{-121/312}\left(-2 q-2 q^2-2 q^4-2 q^6+\dots\right)$ 
  \\
  \hline
  $5$ & $q^{-289/312}\left(-2 q-2 q^3-2 q^4-2 q^5+\dots\right)$ 
  \\
  \hline
  $6$ & $q^{-529/312}\left(-2 q^2-2 q^3-2 q^4-2 q^6+\dots\right)$ 
  \\
  \hline
\end{tabular}
\end{center}
\caption{Small $q$ expansions of the $p=13$ dual $q$-series $X^{(j)}_{(2,3,13)}(q)^\vee$ for $1\leq j \leq 6$, for the Brieskorn sphere $\overline{\Sigma(2, 3, 13)}$.}
\label{tab:mock13-table-match}
\end{table}
Remarkably, these expansions agree  with results of Cheng and Duncan \cite{CD20c} for optimal mock Jacobi forms associated with Brieskorn spheres. See Table 18 on page 55 of \cite{CD20c}. Those listed coefficients match the numerical results  of \cite{ACDGO}, the first few of which are shown in Table \ref{tab:mock13-table-match}, up to a simple reordering of the list of $q$-series. Once again, this correspondence is surprising and suggestive, because the methods for generating these expansions are so very different.

\begin{figure}[h!]
\centering
  \includegraphics[width=0.7\linewidth]{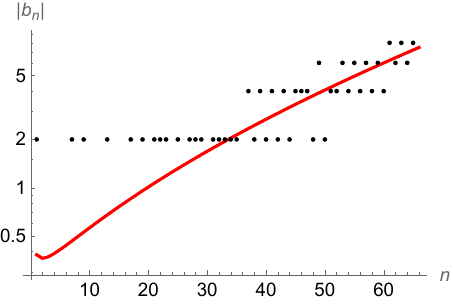}
\caption{This log plot compares the predicted growth for the $q$-series coefficients of $X^{(1)}_{(2,3,13)}(q)^\vee$ given in (\ref{eq:mockgrowth}) for $p=13$ and $j=1$ (red solid line) to the numerically determined coefficients of $X^{(1)}_{(2,3,13)}(q)^\vee$ (black points).}
\label{fig:mock13-growth}
\end{figure}

This makes it interesting to study the large-order growth of the numerical results from \cite{ACDGO} and the optimal mock Jacobi form approach of \cite{CD20c}. The growth data  for this $p=13$ case is shown in Table \ref{tab:p=13secondclass}. Notice the very small value of $\tilde \Delta_{(2,3,13)}=1/312$. This implies extremely slow growth. Indeed, this can be seen in Table \ref{tab:mock13-table-match}, where the first few coefficients are all -2. In fact, one needs to go to very high order, beyond what fits in a table, to find a coefficient -4, and beyond. See Figure \ref{fig:mock13-growth}. It appears that it would require extremely high orders for the leading asymptotic growth rate to set in. This should be compared with the growth for $p=5$, $p=7$ and $p=11$, as shown in Figures \ref{fig:mock5-growth}, \ref{fig:mock7-growth}, and \ref{fig:mock11-growth}. In Figures \ref{fig:mock5-growth} and \ref{fig:mock7-growth}, for  $p=5$ and $p=7$, respectively,  we also see scattered slow growth at low orders, before approaching the asymptotic behavior at higher orders. On the other hand, in Figure \ref{fig:mock11-growth} for $p=11$ we see much more rapid growth, already following the asymptotic form at lower orders.

\begin{figure}[h!]
\centering
  \includegraphics[width=0.65\linewidth]{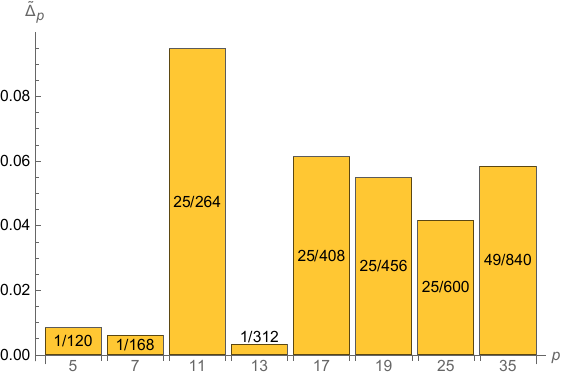}
\caption{Values of $\tilde{\Delta}_{(2,3,p)}$ for the $q$-series associated to orientation-reversed Brieskorn spheres with $p=6k\pm 1$, with $k\in \mathbb Z^+$. Note the wide variation of values. Smaller $\tilde{\Delta}_{(2,3,p)}$ corresponds to slower growth of the $q$-series coefficients.}
\label{fig:histogram2}
\end{figure}

As argued in Section \ref{sec:growth-class1}, the principal determining factor for the growth is the magnitude of $\tilde \Delta_p/\tilde\Delta_{(2,3,p)}$. In Figure \ref{fig:histogram2} we show a histogram of the $\tilde \Delta_{(2,3,p)}$ for the Brieskorn sphere cases with $p=6k\pm 1$. The small magnitude of $\tilde\Delta_{(2,3,13)}$ explains the extremely slow growth. We further see that the growth is indeed much more rapid for $p=17$ and $p=19$, since $\tilde\Delta_{(2,3,17)}$ and $\tilde\Delta_{(2,3,19)}$ are much larger. This also applies to $\tilde\Delta_{(2,3,25)}$ and $\tilde\Delta_{(2,3,35)}$.
\begin{table}[h!]
\begin{tabular}{ | m{3em} | m{12em}| } 
  \hline
  $p=13$ & $X_{(2,3,13)}(q)^\vee$\\ 
  \hline
  $\Delta_{(13,a)}$ & $\left\{ \frac{49}{312},\frac{1}{312},\frac{25}{312},\frac{121}{312},\frac{289}{312},\frac{529}{312}, \right\}$ 
  \\
  \hline
  $\delta_{(13,a)}$ & $\left\{ 1,0,1,1,1,1 \right\}$ 
  \\
  \hline
  $\tilde{b}_{(2,3,13)}$ & $2$ 
  \\
  \hline
  $\widetilde{\Delta}_{(2,3,13)}$ & $1/312$ 
  \\
  \hline
  $c_{\rm eff,\frac{1}{2}}^{T[\overline{M_3}]}$ & $14/13$
  \\
  \hline
\end{tabular}
\caption{Large order growth data, for the growth rate expression \eqref{eq:mockgrowth}, for the dual $q$-series $X^{(j)}_{(2,3,13)}(q)^\vee$, with $j=1, 2, \dots, 6$.} 
\label{tab:p=13secondclass}
\end{table}
\graphicspath{ {./paper plots/} }
\pagebreak
\subsection{Comparisons with other results for Duals for Brieskorn Spheres}
\label{sec:brieskorn-comparisons}
In this section, we compare our results for the effective central charges of the half indices $c_{\rm eff, \frac{1}{2}}^{T[\overline{M_3}]}$ with other results in the literature. Despite being based on completely different methods, we find some correspondences, but interestingly we also find some differences. We first use the results of the previous section to compute the effective central charge for orientation-reversed Brieskorn spheres in the class $\overline{\Sigma(2,3,6k\pm1)}$.
This class contains both known examples, $\overline{\Sigma(2,3,5)}$ and $\overline{\Sigma(2,3,7)}$, as discussed in sections \ref{sec:mock5} and \ref{sec:mock7}, respectively. The first new example in this class,  $\overline{\Sigma(2,3,11)}$, generates dual $q$-series in agreement with a proposal of Zagier \cite{Zag09}, while the second new example,  $\overline{\Sigma(2,3,13)}$, generates dual $q$-series in agreement with {\it optimal} $q$-series studied by Cheng and Duncan \cite{CD20c}. We then turn to another class of Brieskorn spheres, $\overline{\Sigma(s, t, s t\pm 1)}$, studied in \cite{GJ}, which takes an approach based on a conjectured surgery formula \cite{Park21}.
After writing this paper, we shared results for $c_{\rm eff}$ with the authors of \cite{mrunmay}, who also use the conjectured surgery formula, together with modularity and mixed mock modularity arguments, and we comment on the comparison of our results. 
In addition, we use the methods outlined in Appendix B of \cite{ACDGO} to expand our analysis to any orientation-reversed Brieskorn sphere, $\overline{\Sigma(p_1, p_2, p_3)}$, where $p_1, p_2$, and $p_3$ are all mutually coprime. Since we only need the first nonzero term of each $q$-series in the vector, our method can calculate the effective central charge for these manifolds for relatively large values of $p_1$, $p_2$, and $p_3$. For these more general Brieskorn spheres, (\ref{eq:mockgrowth}) takes the form
\begin{eqnarray}
    X^{(j)}_{(p_1,p_2,p_3)}(q)^\vee = \sum\limits_{n=\delta_{(p,j)}}^{\infty} b_n^{(j)} q^n \qquad {\rm with}  \qquad b^{(j)}_n \sim \frac{e^{\pi \sqrt{16\widetilde{\Delta}_{(p_1,p_2,p_3)}\, n}}}{\sqrt{n}},
    \label{eq:mockgrowth2}
\end{eqnarray}
where we omit any constant prefactors. We know that $\tilde{\Delta}_{(p_1,p_2,p_3)}$ takes the form $\frac{m^2}{4p_1p_2p_3}$, where in general $m$ is an integer not divisible by any $p_i$. 
\subsubsection{\boldmath$\overline{\Sigma(2,3,6k\pm1)}$}
\label{sec:236pm1}
In Table \ref{tab:ceff23}, we collect data relevant to the effective central charges coming from $\overline{\Sigma(2,3,6k\pm1)}$.
We agree with the well known results for $k=1$, which are just related to the growth rates of the order $5$ and order $7$ mock theta functions. 
\begin{table}[H]
\begin{tabular}{ | m{3em} | m{4em} | m{4em}| m{4.2em}| m{4.2em}| m{4.2em}| m{4.2em}| m{4.2em}| m{4.2em}| } 
  \hline&
  $\overline{\Sigma(2,3,5)}$ & $\overline{\Sigma(2,3,7)}$ & $\overline{\Sigma(2,3,11)}$ & $\overline{\Sigma(2,3,13)}$ & $\overline{\Sigma(2,3,17)}$ & $\overline{\Sigma(2,3,19)}$ & $\overline{\Sigma(2,3,25)}$ & $\overline{\Sigma(2,3,35)}$  \\ 
  \hline
  $m$ & $1$ & $1$ & $5$ & $1$ & $5$ & $5$ & $5$ & $7$ \\
  \hline
  $c_{\rm eff, \frac{1}{2}}^{T[\overline{M_3}]}$ & $6/5$ & $8/7$ & $36/11$ & $14/13$ & $42/17$ & $44/19$ & $2$  & $12/5$ \\
  \hline
  $c_{\rm eff}$ & $11/5$ & $15/7$ & $47/11$ & $27/13$ & $59/17$ & $63/19$ & $3$ & $17/5$ \\
  \hline
\end{tabular}
\caption{Effective central charge data for $\overline{\Sigma(2,3,6k\pm1)}$, obtained using the resurgent method of \cite{ACDGO}.} 
\label{tab:ceff23}
\end{table}
For $k=2$, for $\overline{\Sigma(2,3,11)}$ we find that the $q$-series results agree with an expression proposed by Zagier in \cite{Zag09} (up to the ordering of the indexing $j$):
\begin{eqnarray}
        M_{11,j}(\tau) = \frac{1}{\eta(\tau)^3}\,\sum\limits_{\substack{m>2|n|/11 \\ n \equiv j \,(\text{mod } 11)}} \left(\frac{-4}{m}\right)\left(\frac{12}{n}\right) \left( m \,\text{sgn}(n) - \frac{n}{6} \right) \, q^{m^2/8 - n^2/264}.
        \label{eq:zagier11}
    \end{eqnarray}
For $j=1, \dots, 5$, expression \eqref{eq:zagier11}  gives:
\begin{align}
    M_{11,1}(q) &= \frac{1}{6} q^{-\frac{1}{264}}\left(5+5 q+15 q^2+60 q^3+125 q^4+313 q^5+620 q^6+1270 q^7+2358 q^8+\dots\right)
    \\
    M_{11,2}(q)&=\frac{1}{6} q^{-\frac{169}{264}}\left( -5 q-22 q^2-60 q^3-155 q^4-357 q^5-781 q^6-1567 q^7-3070 q^8-\dots\right) 
   \\
    M_{11,3}(q)&= \frac{1}{6} q^{-\frac{361}{264}}\left( 5 q^2+33 q^3+99 q^4+268 q^5+605 q^6+1320 q^7+2623 q^8+5104 q^9+ \dots\right)
   \\
   M_{11,4}(q)&= \frac{1}{6} q^{-\frac{49}{264}}\left(-16 q-55 q^2-155 q^3-385 q^4-852 q^5-1816 q^6-3597 q^7-6880 q^8-\dots \right)
   \\
  M_{11,5}(q)&=\frac{1}{6} q^{-\frac{25}{264}}\left(-1+15 q+65 q^2+175 q^3+450 q^4+989 q^5+2105 q^6+4140 q^7+7930 q^8+ \dots \right)
\end{align}
These agree with our $q$-series listed in Table \ref{tab:mock11-table-match}, up to a different labeling convention.

This agreement is surprising, given that the approaches in \cite{Zag09} and in \cite{ACDGO} are so very different, and is suggestive of something deeper.

For $\overline{\Sigma(2,3,13)}$, we find \textit{optimal} $q$-series that appear in the analysis of optimal mock Jacobi functions \cite{CD20c}. As noted in \cite{ACDGO},  the coefficients of our $X_{13}^{(j)}(q)^\vee$ $q$-series, the first terms of which are shown in Table \ref{tab:mock13-table-match}, match those  in Table 18 of \cite{CD20c}. See also Section \ref{sec:mock13}. 

The optimality condition can be rephrased as a statement on the effective central charge, 
\begin{eqnarray}
    \text{optimality} \quad \Longleftrightarrow \quad 
    c_{\rm eff,\frac{1}{2}}^{T[\overline{\Sigma(p_1,p_2,p_3)}]} = 1+\frac{6}{p_1p_2p_3}
\end{eqnarray}
which corresponds to Brieskorn spheres with $\tilde{\Delta}_{(p_1,p_2,p_3)} = \frac{1}{4p_1p_2p_3}$, i.e. $m=1$: see the histogram in Figure \ref{fig:histogram2}. We note that approaches based on the surgery formula in \cite{CCKPG} and \cite{mrunmay} produce $q$-series with different coefficients and analytic structure for $\overline{\Sigma(s,t,kst\pm1)}$ with $k>1$. For this example, the surgery formula identifies dominant singularities of the $q$-series at $q=e^{-2\pi i \frac{2}{5}}$ and $q=e^{-2\pi i \frac{3}{5}}$, a feature present in examples with $k>1$. However, for all of our $q$-series the dominant singularities lie at $q=\pm1$, highlighting a key difference in the methods based on the surgery formula and resurgent continuation. Despite the differences in this example, the asymptotics of the $q$-series obtained from both methods gives the same value for the effective central charge from the orientation-reversed side, $c_{\rm eff,\frac{1}{2}}^{T[\overline{M_3}]}=14/13$.

\subsubsection{\boldmath$\overline{\Sigma(s,t,st\pm1)}$}
\label{sec:ststpm1}
In Tables  \ref{tab:ceffplus} and \ref{tab:ceffminus}, we give the effective central charge data for Brieskorn spheres $\overline{\Sigma(s,t,st\pm1)} = S^3_{1/r}(T_{s,\mp t})$, where $T_{s,t}$ is the $(s,t)$ torus knot. 
\begin{table}[H]
\begin{tabular}{ | m{3em} | m{4em} | m{4em}| m{4.2em}| m{4.2em}| m{4.2em}| m{4.2em}| } 
  \hline&
  $\overline{\Sigma(2,3,7)}$ & $\overline{\Sigma(2,5,11)}$ & $\overline{\Sigma(2,7,15)}$ & $\overline{\Sigma(3,4,13)}$ & $\overline{\Sigma(3,5,16)}$  \\ 
  \hline
  $m$ & $1$ & $3$ & $5$ & $2$ & $4$  \\
  \hline
  $c_{\rm eff, \frac{1}{2}}^{T[\overline{M_3}]}$ & $8/7$ & $82/55$ & $12/7$ & $15/13$ & $7/5$  \\
  \hline
  $c_{\rm eff}$ & $15/7$ & $137/55$ & $19/7$ & $28/13$ & $12/5$   \\
  \hline
\end{tabular}
\caption{Effective central charge data for $\overline{\Sigma(s,t,st+1)}$, obtained using the resurgent method of \cite{ACDGO}.} 
\label{tab:ceffplus} 
\end{table}
\begin{table}[H]
\begin{tabular}{ | m{3em} | m{4em} | m{4em}| m{4.2em}| m{4.2em}| m{4.2em}| } 
  \hline&
  $\overline{\Sigma(2,3,5)}$ & $\overline{\Sigma(2,5,9)}$ & $\overline{\Sigma(2,7,13)}$ & $\overline{\Sigma(3,4,11)}$ & $\overline{\Sigma(3,5,14)}$  \\ 
  \hline
  $m$ & $1$ & $3$ & $3$ & $5$ & $7$  \\
  \hline
  $c_{\rm eff, \frac{1}{2}}^{T[\overline{M_3}]}$ & $6/5$ & $8/5$ & $118/91$ & $47/22$ & $12/5$  \\
  \hline
  $c_{\rm eff}$ & $11/5$ & $13/5$ & $209/91$ & $69/22$ & $17/5$   \\
  \hline
\end{tabular}
\caption{Effective central charge data for $\overline{\Sigma(s,t,st-1)}$, obtained using the resurgent method of \cite{ACDGO}.} 
\label{tab:ceffminus}
\end{table}
For the class $\overline{\Sigma(s,t,st+1)}$, we find that only the results for $\overline{\Sigma(2,3,7)}$, $\overline{\Sigma(2,5,11)}$, and $\overline{\Sigma(2,7,15)}$ match the results proposed in \cite{mrunmay} using the mixed modularity approach. Our $c_{\rm eff}$ values do not agree with those in \cite{mrunmay} for all other Brieskorn spheres analyzed in this class.
The paper \cite{mrunmay} finds values of $m$ that are irrational, and thus are not associated to a Chern-Simons value. 

For the class $\overline{\Sigma(s,t,st-1)}$, a description in terms of the mixed modularity of the $\widehat{Z}$ invariants is unavailable and thus we may only compare to numerical estimates made in \cite{mrunmay} using the surgery formula. In general, our results are not consistent with their estimates. 
These comparisons can be found in Figure 11 of \cite{mrunmay}. 
Once again, given the very different approaches in this paper and in \cite{mrunmay}, it will be important to understand the comparisons. In particular, in our approach based on resurgent preservation of relations, there appears to be no fundamental difference between the cases $\overline{\Sigma(s,t,st+1)}$ and $\overline{\Sigma(s,t,st-1)}$, whereas these cases are significantly different in the mixed modularity approach.

\subsubsection{\textbf{Other Brieskorn Spheres}}
Our resurgent method, and accompanying numerical algorithm, are not limited to the cases discussed in the previous subsections.
In Table \ref{tab:ceffother} we give results for a selection of orientation-reversed Brieskorn spheres that fall outside of these previous classes. 
\begin{table}[H]
\begin{tabular}{ | m{3em} | m{4em} | m{4em}| m{4.2em}| m{4.2em}| m{4.2em}| } 
  \hline&
  $\overline{\Sigma(2,5,7)}$ & $\overline{\Sigma(3,4,5)}$ & $\overline{\Sigma(3,5,7)}$ & $\overline{\Sigma(3,5,8)}$ & $\overline{\Sigma(4,5,7)}$  \\ 
  \hline
  $m$ & $1$ & $1$ & $2$ & $2$ & $2$  \\
  \hline
  $c_{\rm eff, \frac{1}{2}}^{T[\overline{M_3}]}$ & $38/35$ & $11/10$ & $43/35$ & $6/5$ & $41/35$  \\
  \hline
  $c_{\rm eff}$ & $73/35$ & $21/10$ & $78/35$ & $11/5$ & $76/35$   \\
  \hline
\end{tabular}
\caption{Effective central charge data for Brieskorn spheres $\overline{\Sigma(p_1,p_2,p_3)}$ not belonging to the previous classes, obtained using the resurgent method of \cite{ACDGO}.} 
\label{tab:ceffother}
\end{table}

\subsubsection{\textbf{Further comparisons}}
There is also an interesting connection between the results obtained in this work and results presented in \cite{DMZ12}. Appendix A.3 of \cite{DMZ12} contains tables of coefficients of $q$-series associated with special weight 1 mock Jacobi forms $\mathcal{Q}_M$, along with the values of the minimal discriminants $\Delta_{\min}$ of the vector valued forms. For their examples with $M=30,42,66,78,102,114,138$, the set of $q$-series associated with $\mathcal{Q}_M$ match the set of dual $q$-series for the orientation reversed Brieskorn spheres $\overline{\Sigma(2, 3, p)}$, with $M=6p$,
found using the algorithm of \cite{ACDGO}, and the corresponding $\Delta_{\min}$ values in \cite{DMZ12} agree with our values of $-m^2$. We have also confirmed the same agreement with \cite{DMZ12} concerning dual $q$-series and $\Delta_{\min}$ for $\overline{\Sigma(2, 5, 7)}$, $\overline{\Sigma(3, 5, 7)}$, $\overline{\Sigma(2, 5,11)}$, and $\overline{\Sigma(2, 5, 13)}$, corresponding to $M=70,105,110$, and $130$ in Appendix A.3 of \cite{DMZ12}. The physics behind this correspondence will be explored in future work.

\section{Conclusions}

In this paper we have shown that the new methods introduced  in \cite{CDGG,ACDGO} for continuing Chern-Simons partition functions across the natural boundary produce integer-coefficient $q$-series with growth rate that is determined by the algebraic structure of the resurgent cyclic orbits, combined with knowledge of the power of the leading terms of the $q$-series. This permits the direct evaluation of the effective central charge $c_{\rm eff}$. This can be done for the duals of individual false theta functions, as well as for the particular linear combinations of false theta functions that characterize Chern-Simons theory on Brieskorn spheres.
The algebraic structure of the resurgent cyclic orbits is encoded in the Mordell-Borel integrals, evaluated on the Stokes line, and the leading term is determined numerically by imposing the fundamental resurgent principle of preservation of relations. This is a new and more stringent test of resurgence of the Chern-Simons path integral. Furthermore, the resurgent continuation method produces results for a much wider class of manifolds on which the Chern-Simons theory is defined, and we present results for other Brieskorn spheres that do not appear to be computable with any other current method.

We compare our results with other approaches to computing $c_{\rm eff}$, revealing intriguing correspondences and differences, which deserve future study. 
The first two new Brieskorn sphere examples in the class  $\overline{\Sigma(2,3,6k\pm 1)}$, $\overline{\Sigma(2,3,11)}$ and $\overline{\Sigma(2,3,13)}$, generate dual $q$-series in agreement with a proposal of Zagier \cite{Zag09}, and  {\it optimal} $q$-series of Cheng and Duncan \cite{CD20c}, respectively. There are also many $q$-series in common with those appearing in  \cite{DMZ12}, related to weight 1 mock Jacobi forms. Therefore the growth rates and $c_{\rm eff}$ also agree. This is surprising, given the very different methods used in those papers.
For other cases, we compare our results with a companion paper \cite{mrunmay}, based on an approach utilizing the surgery formula proposed in \cite{Park21}, combined with mixed mock modularity arguments. Our approach does not rely on this conjectured surgery formula. 
Interestingly, the results in \cite{mrunmay} agree for certain examples, and differ in other examples from our resurgent analytic continuation method.
It is possible that a certain regularization of the surgery formula, or utilizing the full $SL(2,\mathbb{Z})$ structure in the surgery formula, may provide an explanation for these differences, and which we leave for future work. For resurgent analysis, it is intimately related to understanding precise conditions for uniqueness, one of the central topics in this paper and in \cite{CDGG,ACDGO}.

\section{Acknowledgements}

Thanks to John Chae, Miranda C. N. Cheng, Daniele Dorigoni, Jean \'Ecalle, Shimal Harichurn, Mrunmay Jagadale, Albrecht Klemm, Dmitry Noshchenko, Davide Passaro, and Don Zagier for discussions, ideas, help, advice, support and inspiration that have greatly benefited this project. We are especially grateful to Shimal Harichurn, Mrunmay Jagadale, Dmitry Noshchenko and Davide Passaro for sharing their results in \cite{mrunmay}.
The work of OC is supported in part by the U.S. National Science Foundation, Division of Mathematical Sciences, Award NSF DMS-2206241. The work of GD and GA is supported in part by the U.S. Department of Energy, Office of High Energy Physics, Award DE-SC0010339.
The work of SG was supported in part by a Simons Collaboration Grant on New Structures in Low-Dimensional Topology, by the NSF grant DMS-2245099, and by the U.S. Department of Energy, Office of Science, Office of High Energy Physics, under Award No. DE-SC0011632.
The authors OC, GD and SG would like to thank 
the Galileo Galilei Institute for Theoretical Physics for hospitality, and the INFN for partial support, during the workshop “Resurgence and Modularity in QFT and String Theory”, Spring 2024. GD thanks the Max Planck Institute for Mathematics, Bonn, for support during the program "Combinatorics, Resurgence and Algebraic Geometry in Quantum Field Theory", August 2024. GA and O\"O thank L'\'Ecole de Physique des Houches for support during the 2024 Summer School "Quantum Geometry".

\appendix

\numberwithin{equation}{section}

\section{Data for Dual of False Thetas}
\label{sec:results-p-class}

\begin{table}[h!]
\begin{center}
\begin{tabular}{ | m{3em} | m{15em}| } 
  \hline
  $p=11$  & $\Psi_{11}$ \\ 
  \hline
  $\Delta_{(11,a)}$ & $\left\{ \frac{1}{88},\frac{1}{11},\frac{9}{88},\frac{4}{11},\frac{25}{88},\frac{9}{11},\frac{49}{88},\frac{16}{11},\frac{81}{88},\frac{25}{11} \right\}$ 
  \\
  \hline
  $\delta_{(11,a)}$ & $\left\{ 0,1,0,1,1,1,1,4,1,3 \right\}$ 
  \\
  \hline
  $\widetilde{\Delta}_{11}$ & $9/44$ 
  \\
  \hline
  $\tilde{b}_{11}$ & $1/2$ 
  \\
  \hline
\end{tabular}
\end{center}
\caption{Large order growth data for the dual $q$-series $\Psi^{(2j-1)}_{11} \left(i \sqrt{q}\right)^\vee$ and $\Psi^{(2j)}_{11} \left(-q\right)^\vee$. This data determines the large order growth of the coefficients in the $q$-series via the expressions in \eqref{eq:p-growth-odd}-\eqref{eq:p-growth-even}.} 
\label{tab:p=11firstclass}
\end{table}
\begin{table}[h!]
\begin{center}
\begin{tabular}{ | m{3em} | m{20em}| } 
  \hline
  $p=13$ & $\Psi_{13}^\vee$ \\ 
  \hline
  $\Delta_{(13,a)}$ & $\left\{ \frac{1}{104},\frac{1}{13},\frac{9}{104},\frac{4}{13},\frac{25}{102},\frac{9}{13},\frac{49}{104},\frac{16}{13},\frac{81}{104},\frac{25}{13},\frac{121}{104},\frac{36}{13} \right\}$ 
  \\
  \hline
  $\delta_{(13,a)}$ & $\left\{ 1,1,0,1,1,1,3,3,1,2,2,3 \right\}$ 
  \\
  \hline
  $\widetilde{\Delta}_{13}$ & $9/52$ 
  \\
  \hline
  $\tilde{b}_{13}$ & $1$ 
  \\
  \hline
\end{tabular}
\end{center}
\caption{Large order growth data for the dual $q$-series $\Psi^{(2j-1)}_{13} \left(i \sqrt{q}\right)^\vee$ and $\Psi^{(2j)}_{13} \left(-q\right)^\vee$. This data determines the large order growth of the coefficients in the $q$-series via the expressions in \eqref{eq:p-growth-odd}-\eqref{eq:p-growth-even}.} 
\label{tab:p=13firstclass}
\end{table}
\begin{table}[h!]
\begin{center}
\begin{tabular}{ | m{3em} | m{24em}| } 
  \hline
  $p=15$ & $\Psi_{15}^\vee$ \\ 
  \hline
  $\Delta_{(15,a)}$ & $\left\{ \frac{1}{120},\frac{1}{15},\frac{9}{120},\frac{4}{15},\frac{25}{120},\frac{9}{15},\frac{49}{120},\frac{16}{15},\frac{81}{120},\frac{25}{15},\frac{121}{120},\frac{36}{15},\frac{169}{120},\frac{49}{15} \right\}$ 
  \\
  \hline
  $\delta_{(15,a)}$ & $\left\{ 0,1,0,1,0,1,1,2,1,2,1,4,2,4 \right\}$ 
  \\
  \hline
  $\widetilde{\Delta}_{15}$ & $25/60$ 
  \\
  \hline
  $\tilde{b}_{15}$ & $1/4$ 
  \\
  \hline
\end{tabular}
\end{center}
\caption{Large order growth data for the dual $q$-series $\Psi^{(2j-1)}_{15} \left(i \sqrt{q}\right)^\vee$ and $\Psi^{(2j)}_{15} \left(-q\right)^\vee$. This data determines the large order growth of the coefficients in the $q$-series via the expressions in \eqref{eq:p-growth-odd}-\eqref{eq:p-growth-even}.} 
\label{tab:p=15firstclass}
\end{table}
\begin{table}[h!]
\begin{center}
\begin{tabular}{ | m{3em} | m{27em}| } 
  \hline
  $p=17$ & $\Psi_{17}^\vee$ \\ 
  \hline
  $\Delta_{(17,a)}$ & $\left\{ \frac{1}{136},\frac{1}{17},\frac{9}{136},\frac{4}{17},\frac{25}{136},\frac{9}{17},\frac{49}{136},\frac{16}{17},\frac{81}{136},\frac{25}{17},\frac{121}{136},\frac{36}{17},\frac{169}{136},\frac{49}{17},\frac{225}{136},\frac{64}{17} \right\}$ 
  \\
  \hline
  $\delta_{(17,a)}$ & $\left\{ 0,1,0,1,1,1,3,1,1,2,1,5,3,4,2,4 \right\}$ 
  \\
  \hline
  $\widetilde{\Delta}_{17}$ & $9/68$ 
  \\
  \hline
  $\tilde{b}_{17}$ & $1$ 
  \\
  \hline
\end{tabular}
\end{center}
\caption{Large order growth data for the dual $q$-series $\Psi^{(2j-1)}_{17} \left(i \sqrt{q}\right)^\vee$ and $\Psi^{(2j)}_{17} \left(-q\right)^\vee$. This data determines the large order growth of the coefficients in the $q$-series via the expressions in \eqref{eq:p-growth-odd}-\eqref{eq:p-growth-even}.} 
\label{tab:p=17firstclass}
\end{table}
\begin{table}[H]
\begin{tabular}{ | m{3em} | m{30em}| } 
  \hline
  $p=19$ & $\Psi_{19}^\vee$\\ 
  \hline
  $\Delta_{(19,a)}$ & $\left\{ \frac{1}{152},\frac{1}{19},\frac{9}{152},\frac{4}{19},\frac{25}{152},\frac{9}{19},\frac{49}{152},\frac{16}{19},\frac{81}{152},\frac{25}{19},\frac{121}{152},\frac{36}{19},\frac{169}{152},\frac{49}{19},\frac{225}{152},\frac{64}{19},\frac{289}{152},\frac{81}{19} \right\}$ 
  \\
  \hline
  $\delta_{(19,a)}$ & $\left\{ 0,1,0,1,0,1,1,1,1,2,1,2,1,3,2,4,2,5 \right\}$ 
  \\
  \hline
  $\widetilde{\Delta}_{19}$ & $25/76$ 
  \\
  \hline
  $\tilde{b}_{19}$ & $1/8$ 
  \\
  \hline
\end{tabular}
\caption{Large order growth data for the dual $q$-series $\Psi^{(2j-1)}_{19} \left(i \sqrt{q}\right)^\vee$ and $\Psi^{(2j)}_{19} \left(-q\right)^\vee$. This data determines the large order growth of the coefficients in the $q$-series via the expressions in \eqref{eq:p-growth-odd}-\eqref{eq:p-growth-even}.} 
\label{tab:p=19firstclass}
\end{table}

\pagebreak

\section{Data for Duals for Orientation-Reversed Brieskorn Spheres}
\label{sec:results-mock-class}
\begin{table}[h!]
\begin{tabular}{ | m{3em} | m{16em}| } 
  \hline
  $p=17$ & $X_{17}^\vee$ \\ 
  \hline
  $\Delta_{(17,a)}$ & $\left\{ \frac{121}{408},\frac{25}{408},\frac{1}{408},\frac{49}{408},\frac{169}{408},\frac{361}{408},\frac{625}{408},\frac{961}{408} \right\}$ 
  \\
  \hline
  $\delta_{(17,a)}$ & $\left\{ 1,0,0,1,1,1,2,3 \right\}$ 
  \\
  \hline
  $\tilde{b}_{(2,3,17)}$ & $1/3$ 
  \\
  \hline
  $\widetilde{\Delta}_{(2,3,17)}$ & $25/408$
  \\
  \hline
\end{tabular}
\caption{Large order growth data, for the growth rate expression \eqref{eq:mockgrowth}, for the dual $q$-series $X^{(j)}_{(2,3,17)}(q)^\vee$, with $j=1, 2, \dots, 8$.} 
\label{tab:p=17secondclass}
\end{table}

\begin{table}[h!]
\begin{tabular}{ | m{3em} | m{18em}| } 
  \hline
  $p=19$ & $X_{19}^\vee$ \\ 
  \hline
  $\Delta_{(19,a)}$ & $\left\{ \frac{169}{456},\frac{49}{456},\frac{1}{456},\frac{25}{456},\frac{121}{456},\frac{289}{456},\frac{529}{456},\frac{841}{456},\frac{1225}{456} \right\}$ 
  \\
  \hline
  $\delta_{(19,a)}$ & $\left\{ 1,1,0,0,1,1,2,2,3 \right\}$ 
  \\
  \hline
  $\tilde{b}_{(2,3,19)}$ & $1/2$ 
  \\
  \hline
  $\widetilde{\Delta}_{(2,3,19)}$ & $25/456$ 
  \\
  \hline
\end{tabular}
\caption{Large order growth data, for the growth rate expression \eqref{eq:mockgrowth}, for the dual $q$-series $X^{(j)}_{(2,3,19)}(q)^\vee$, with $j=1, 2, \dots, 9$.} 
\label{tab:p=19secondclass}
\end{table}

\begin{table}[h!]
\begin{tabular}{ | m{3em} | m{14em}| } 
  \hline
  $\!\!\overline{\Sigma(2,5,7)}$ & $X_{(2,5,7)}(q)^\vee$ \\ 
  \hline
  $\Delta_{(2,5,7)}$ & $\left\{ \frac{1}{280},\frac{9}{280},\frac{81}{280},\frac{121}{280},\frac{169}{280},\frac{529}{280}\right\}$ 
  \\
  \hline
  $\delta_{(2,5,7)}$ & $\left\{ 0,2,1,1,1,2 \right\}$ 
  \\
  \hline
  $\tilde{b}_{(2,5,7)}$ & $-2$ 
  \\
  \hline
  $\widetilde{\Delta}_{(2,5,7)}$ & $1/280$ 
  \\
  \hline
\end{tabular}
\caption{Large order growth data, for the growth rate expression \eqref{eq:mockgrowth} (with the appropriate matrix $M$ constructed in Section 12 of \cite{ACDGO}), for the dual $q$-series $X^{(j)}_{(2,5,7)}(q)^\vee$, with $j=1, 2, \dots, 6$.} 
\label{tab:p=257}
\end{table}

\begin{table}[h!]
\begin{tabular}{ | m{3.2em} | m{14em}| } 
  \hline
  $\!\!\overline{\Sigma(3,4,5)}$ & $X_{(3,4,5)}(q)^\vee$ \\ 
  \hline
  $\Delta_{(3,4,5)}$ & $\left\{ \frac{1}{240},\frac{4}{240},\frac{49}{240},\frac{121}{240},\frac{169}{240},\frac{196}{240}\right\}$ 
  \\
  \hline
  $\delta_{(3,4,5)}$ & $\left\{ 0,1,2,1,1,1 \right\}$ 
  \\
  \hline
  $\tilde{b}_{(3,4,5)}$ & $-2$ 
  \\
  \hline
  $\widetilde{\Delta}_{(3,4,5)}$ & $1/240$ 
  \\
  \hline
\end{tabular}
\caption{Large order growth data, for the growth rate expression \eqref{eq:mockgrowth} (with the appropriate matrix $M$ constructed in Section 12 of \cite{ACDGO}), for the dual $q$-series $X^{(j)}_{(3,4,5)}(q)^\vee$, with $j=1, 2, \dots, 6$.} 
\label{tab:p=345}
\end{table}

\begin{table}[H]
\begin{center}
\begin{tabular}{ | m{5em} | m{20em}| } 
  \hline
  $\overline{\Sigma(2,5,11)}$  & $X_{(2,5,11)}(q)^\vee$ \\ 
  \hline
  $\Delta_{(2,5,11)}$ & $\left\{\frac{1}{440},\frac{9}{440},\frac{49}{440},\frac{81}{440},\frac{169}{440},\frac{289}{440},\frac{361}{440},\frac{529}{440},\frac{841}{440},\frac{1521}{440}   \right\}$ 
  \\
  \hline
  $\delta^{(2,5,11)}_{j}$ & $\left\{ 0,0,1,1,1,1,1,2,2,4 \right\}$ 
  \\
  \hline
  $\tilde{b}_{(2,5,11)}$ & $2/3$ 
  \\
  \hline
  $\widetilde{\Delta}_{(2,5,11)}$ & $9/440$ 
  \\
  \hline
\end{tabular}
\end{center}
\caption{Large order growth data, for the growth rate expression \eqref{eq:mockgrowth}, for the dual $q$-series $X^{(j)}_{(2,5,11)}(q)^\vee$, with $j=1, 2, \dots, 10$.} 
\label{tab:2511growthdata}
\end{table}
\begin{table}[H]
\begin{center}
\begin{tabular}{ | m{5em} | m{36em}| } 
  \hline
  $\overline{\Sigma(3,4,13)}$  & $X_{(3,4,13)}(q)^\vee$ \\ 
  \hline
  $\Delta_{(3,4,13)}$ & $\left\{\frac{1}{624},\frac{4}{624},\frac{25}{624},\frac{49}{624},\frac{100}{624},\frac{121}{624},\frac{196}{624},\frac{289}{624},\frac{361}{624},\frac{484}{624},\frac{529}{624},\frac{841}{624},\frac{961}{624},\frac{1156}{624},\frac{1225}{624},\frac{1681}{624},\frac{2116}{624},\frac{2809}{624}  \right\}$ 
  \\
  \hline
  $\delta^{(3,4,13)}_{j}$ & $\left\{ 0,0,4,11,2,1,2,2,1,2,1,2,2,2,2,3,4,6 \right\}$ 
  \\
  \hline
  $\tilde{b}_{(3,4,13)}$ & $-2$ 
  \\
  \hline
  $\widetilde{\Delta}_{(3,4,13)}$ & $4/624$ 
  \\
  \hline
\end{tabular}
\end{center}
\caption{Large order growth data, for the growth rate expression \eqref{eq:mockgrowth}, for the dual $q$-series $X^{(j)}_{(3,4,13)}(q)^\vee$, with $j=1, 2, \dots, 18$.} 
\label{tab:3413growthdata}
\end{table}
\begin{table}[H]
\begin{center}
\begin{tabular}{ | m{5em} | m{25em}| } 
  \hline
  $\overline{\Sigma(2,7,15)}$  & $X_{(2,7,15)}(q)^\vee$ \\ 
  \hline
  $\Delta_{(2,7,15)}$ & $\{\frac{1}{840},\frac{9}{840},\frac{25}{840},\frac{81}{840},\frac{121}{840},\frac{169}{840},\frac{289}{840},\frac{361}{840},\frac{529}{840},\frac{625}{840},\frac{729}{840},\frac{961}{840},$ 
  \\
  & $\frac{1089}{840},\frac{1521}{840},\frac{1681}{840},\frac{2209}{840},\frac{2809}{840},\frac{3025}{840},\frac{3721}{840},\frac{4761}{840},\frac{6889}{840} \}$ 
  \\
  \hline
  $\delta_{(2,7,15)}$ & $\left\{ 0,0,0,1,1,1,1,1,1,1,1,2,2,2,2,3,4,5,5,6,9 \right\}$ 
  \\
  \hline
  $\tilde{b}_{(2,7,15)}$ & $-1/2$ 
  \\
  \hline
  $\widetilde{\Delta}_{(2,7,15)}$ & $25/840$ 
  \\
  \hline
\end{tabular}
\end{center}
\caption{Large order growth data, for the growth rate expression \eqref{eq:mockgrowth}, for the dual $q$-series $X^{(j)}_{(2,7,15)}(q)^\vee$, with $j=1, 2, \dots, 21$.} 
\label{tab:2715growthdata}
\end{table}

\pagebreak

\section{Subleading growth}
\label{sec:sub}

In this Appendix we outline how to extend the saddle point analysis in Section \ref{sec:legendre} to incorporate subleading corrections. The simplest correction amounts to a shift of the order $n$ to incorporate the exponent of $\qt^{\, -\alpha}$:
\begin{eqnarray}
 b_n \sim \frac{e^{2\pi \sqrt{c\, n}}}{\sqrt{n}} \quad \longrightarrow \quad b_n \sim \frac{e^{2\pi \sqrt{c\,(n-\alpha)}}}{\sqrt{n-\alpha}}.
 \label{eq:shift}
\end{eqnarray}
For the duals of individual false theta functions, we have the following expression for the shift parameter $\alpha$ in \eqref{eq:shift},
\begin{eqnarray}
    \alpha=\begin{cases}
    \frac{a^2}{8p}, \quad a \text{ odd} \\
    \frac{a^2}{4p}, \quad a \text{ even}
\end{cases},
\label{eq:alpha-p}
\end{eqnarray}
and for the duals related to orientation-reversed Brieskorn spheres $\overline{\Sigma(2,3,p)}$:
\begin{eqnarray}
    \alpha = \frac{(6j-p)^2}{24p}
    \label{eq:alpha-mock}
\end{eqnarray}
The subleading correction form in \eqref{eq:shift} was investigated long ago for the order $3$ mock theta functions \cite{Dra}:
\begin{eqnarray}
    \Phi^{(1)}_3(i \sqrt{q})^\vee &=& \frac{1}{2} f(-q)
    \longrightarrow 
    \quad b_n \sim -\frac{1}{4}\frac{e^{\pi \sqrt{(n-1/24)/6}}}{\sqrt{n-1/24}}
    \\
    \Phi^{(2)}_3(-q)^\vee &=& -q \, \omega(q) \longrightarrow \quad b_n \sim -\frac{1}{4}\frac{e^{\pi \sqrt{(n-1/3)/3}}}{\sqrt{n-1/3}}  
\end{eqnarray}
This matches our general result \eqref{eq:alpha-p} for $p=3$, and $a=1, 2$, with the identification: $\Phi_3^{(1)}(i\sqrt{q})^\vee =\frac{1}{2} f(q)$, and $\Phi^{(2)}_3(-q)^\vee = -q \, \omega(q)$. 

For the mock theta functions of order $10$ we have the following subleading corrections to the large-order growth:
\begin{align}
    \Phi^{(1)}_5(i\sqrt{q})^\vee = X(-q) &
    \quad\longrightarrow\quad  b_n \sim \sin\left(\frac{\pi}{5}\right)\frac{e^{\pi \sqrt{(n-1/40)/10}}}{\sqrt{5(n-1/40)}}, \\
    \Phi^{(2)}_5(-q)^\vee = - \psi(q) &
    \quad\longrightarrow\quad b_n \sim -\sin\left(\frac{\pi}{5}\right)\frac{e^{\pi \sqrt{(n-1/5)/5}}}{\sqrt{5(n-1/5)}}, \\
    \Phi^{(3)}_5(i\sqrt{q})^\vee = \chi(-q)&
    \quad\longrightarrow\quad b_n \sim -\sin\left(\frac{2\pi}{5}\right)\frac{e^{\pi \sqrt{(n-9/40)/10}}}{\sqrt{5(n-9/40)}}, \\
    \Phi^{(4)}_5(-q)^\vee = -q \, \phi(q) &
    \quad\longrightarrow\quad b_n \sim -\sin\left(\frac{2\pi}{5}\right)\frac{e^{\pi \sqrt{(n-4/5)/5}}}{\sqrt{5(n-4/5)}}. 
\end{align}
\begin{figure}[h!]
\centering
  \includegraphics[width=0.45\linewidth]{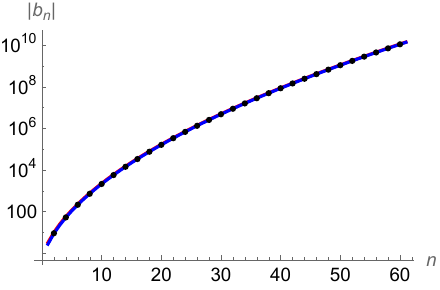}
  \includegraphics[width=0.45\linewidth]{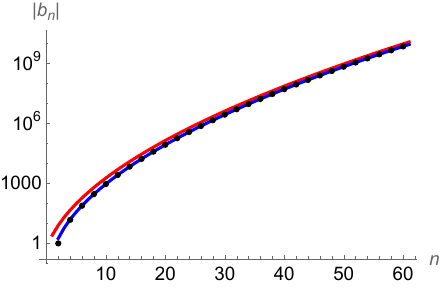}
\caption{Comparison plots including subleading growth of $\Psi^{(a)}_7(-q)^\vee$ for $a=2$ (left) and $a=6$ (right). In both figures, we plot the estimated growth without subleading corrections as a solid red line, and the estimated growth with subleading corrections as a solid blue line. Then, we plot every other coefficient of $\Psi^{(a)}_{7}(-q)^\vee$ as black points. We see that subleading corrections are more pronounced for larger values of $a$.}
\label{fig:p7subgrowth}
\end{figure}
\begin{figure}[h!]
\centering
  \includegraphics[width=0.45\linewidth]{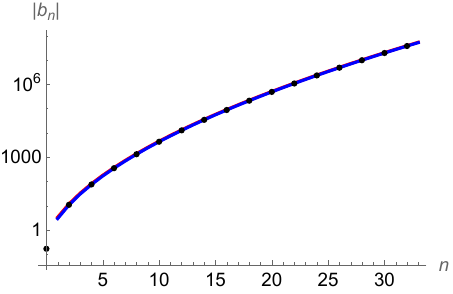}
  \includegraphics[width=0.45\linewidth]{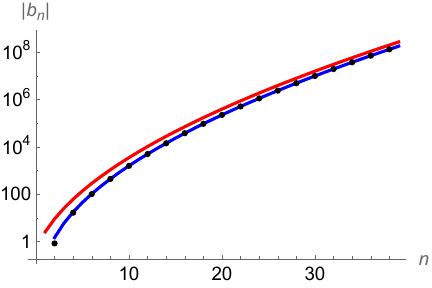}
\caption{Comparison plots including subleading growth of $X^{(j)}_{(2,3,11)}(q)^{\vee}$ for $j=1$ (left) and $j=5$ (right). In both figures, we plot the estimated growth without subleading corrections as a solid red line, and the estimated growth with subleading corrections as a solid blue line. Then, we plot every other coefficient (to avoid clutter) of $X^{(j)}_{(2,3,11)}(q)^\vee$ as black points. We see that subleading corrections are more pronounced for larger values of $j$.}
\label{fig:m11subgrowth}
\end{figure}
From \eqref{eq:alpha-p}-\eqref{eq:alpha-mock}, we see that the effect of the subleading correction is more pronounced for larger values of $p$ and $a$ for the first resurgent orbit, and for larger values of $p$ and $j$ for the second resurgent orbit. 
In Figure \ref{fig:p7subgrowth} we show the effect for $\Psi_7^{(a)\, \vee}$ for $a=2$ [left] and $a=6$
[right]. In Figure \ref{fig:m11subgrowth} we show the effect for $X_{(2,3,11)}^{(j)\, ^\vee}$ for $j=1$ [left] and $j=5$
[right]. In both cases, the correction is not visible in the left-hand plots, with small $a$ and $j$, but becomes noticeable in the right-hand plots, with larger $a$ or $j$.

\begin{figure}[H]
\centering
  \includegraphics[width=0.6\linewidth]{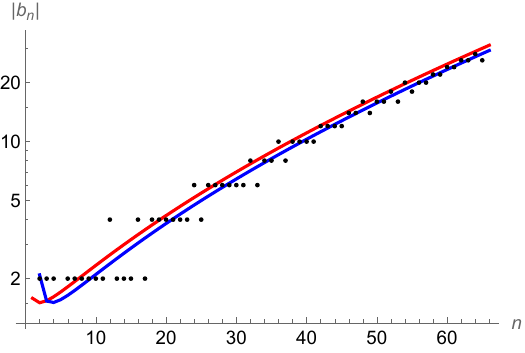}
\caption{The growth is very slow for $X^{(6)}_{(2,3,13)}(q)^{\vee}$, since the $\tilde\Delta_{(2,3,13)}$ parameter is small: see Figure \ref{fig:histogram2}. The red curve shows the leading growth expression, while the blue curve shows the inclusion of the subleading correction \eqref{eq:shift}, with the parameter $\alpha$ given by \eqref{eq:alpha-mock}. The subleading expression is noticeably better.}
\label{fig:m13subgrowth}
\end{figure}

In Figure \ref{fig:m13subgrowth} we show the effect of the subleading correction for $X_{(2,3,13)}^{(6)}(q)^\vee$, which is an interesting case since the growth rate of the coefficients is very slow, as can be seen from the histogram in Figure \ref{fig:histogram2}. We see that the subleading correction plot fits better the slow growth rate.

\end{document}